\documentclass[twocolumn]{aastex631}
\usepackage{float}
\usepackage{amsmath}
\usepackage{rotating}
\usepackage{amssymb}
\usepackage{hyperref}
\usepackage[normalem]{ulem}
\usepackage{graphicx}
\usepackage{xcolor}

\def\gtrsim{\mathrel{\hbox{\rlap{\hbox{\lower4pt\hbox{$\sim$}}}\hbox{$>$}}}}
\def\lesssim{\mathrel{\hbox{\rlap{\hbox{\lower4pt\hbox{$\sim$}}}\hbox{$<$}}}}

\def\farcm{\hbox{$.\!\!^{\prime}$}}

\newcommand {\nustar} {\textsl{NuSTAR}}
\newcommand {\suzaku} {\textsl{Suzaku}}

\def \src {LS\,5039}

\begin{document}

\title{NuSTAR observation of LS 5039}

\author{Igor Volkov}
\affiliation{The George Washington University, Department of Physics, 725 21st St NW, Washington, DC  20052}
\author{Oleg Kargaltsev}
\affiliation{The George Washington University, Department of Physics, 725 21st St NW, Washington, DC  20052}
\author{George  Younes}
\affiliation{The George Washington University, Department of Physics, 725 21st St NW, Washington, DC  20052}
\author{Jeremy Hare}
\altaffiliation{NASA Postdoctoral Program Fellow}
\affiliation{NASA Goddard Space Flight Center, Greenbelt MD, 20771, USA}
\author{George Pavlov}
\affiliation{Pennsylvania State University, Department of Astronomy \& Astrophysics, 525 Davey Laboratory, University Park, PA 16802}

\email{kargaltsev@email.gwu.edu}

\begin{abstract}

LS 5039 is a 
 high-mass $\gamma$-ray
binary 
 hosting a compact object of unknown type. 
{\sl NuSTAR} 
observed LS 5039 during the entire 3.9 day binary
period.
We performed  a periodic signal  search up to 1000 Hz which did not produce 
credible period candidates.
We do see the 9.05 s period candidate, originally reported by 
\cite{Yoneda20} using the same data, in  the  Fourier  power spectrum, but we find that the statistical significance of this feature is too low to claim it as a real detection.  
We also did not find 
significant bursts or quasi-periodic variability.  The modulation with the orbital period is clearly seen and remains unchanged over a decade timescale when compared to the earlier 
 {\sl Suzaku} light curve. The joint analysis of the \nustar\ and \suzaku\ XIS data shows that the 0.7--70 keV spectrum can be satisfactory described by a single absorbed power-law model  with no evidence of cutoff at higher energies. The slope of the spectrum anti-correlates with the flux during the binary orbit.  Therefore,  if LS 5039 hosts a young neutron star,
its X-ray pulsations appear to be 
outshined by the intrabinary shock emission. The lack of spectral 
lines and/or an 
exponential cutoff at higher energies suggests that the putative neutron star is not actively accreting.
Although a black hole scenario still remains a possibility, the lack of variability or Fe K$\alpha$ lines, which typically accompany accretion, makes it less likely.

\end{abstract}


\section{INTRODUCTION}

High-mass $\gamma$-ray
binaries (HMGBs) consist of a compact object, either a neutron star (NS) or black  hole (BH), orbiting a
hot, massive O or B type star.   
The number of known HMGBs 
has been increasing
in recent years \citep{2019arXiv190103624P}. However, in all but two HMGBs (LS\,2883/B1259--63 and MT91\,213/TeV J2031+4130), 
which host young pulsars, the nature of the compact object remains unknown. 
 Two scenarios
for 
high-energy emission of HMGBs
 are 
 usually discussed: (1) interaction of the
pulsar wind with the wind of the massive donor star leading to the 
formation of an intrabinary shock, or (2) jets powered by accretion
onto a compact object (likely a BH). All HMGBs that have been observed with VLBI exhibit extended radio  
emission on milliarcsecond scales, which can be attributed to either a pulsar-wind nebula  or to 
jets 
produced by accretion onto a BH (i.e., the classical microquasar scenario).

X-ray observations can provide important information for 
 understanding the nature of HMGBs. Therefore, 
 these objects have been extensively observed in soft and hard X-rays. 
  In particular, {\sl NuSTAR} observed LS \,2883 \citep{2015MNRAS.454.1358C}, MT91\,213 \citep{2017ApJ...843...85L}, 1FGL\,J1018.6--5856  \citep{2015ApJ...806..166A},  HESS J0632+057 \citep{2019arXiv190803083P,2020ApJ...888..115A}, LS\,I\,+61$^\circ$303 \citep{2020MNRAS.tmp.2539M}, and LMC P3 \citep{2020AAS...23545701C}.  All of these HMGBs were also observed in soft X-rays with {\sl XMM-Newton}, {\sl Swift}, {\sl Suzaku}, and {\sl Chandra} (see e.g., \citealt{2009ApJ...697..592T,2013A&ARv..21...64D,2014AN....335..301K}). In all cases, the X-ray spectra appear to be consistent with featureless power-laws (PLs) with photon indices $\Gamma\simeq1.3-2$. 
  Joint fits to the soft X-ray and hard X-ray spectra provide no evidence of a spectral cut-off in the {\sl NuSTAR} band.   No periodicity associated with the compact object spin has been found in any of the X-ray data (including two systems with known radio pulsars -- LS 2883 with PSR B1259--63 and MT91 213 with PSR J2032+4127). Using NuSTAR data, 
  \cite{2015ApJ...806..166A} 
   and \cite{2020ApJ...888..115A}
   found no short-term variability, quasi-periodic oscillations, red noise, or any other temporal or spectral evidence of accretion in 1FGL J1018.6--5856 and HESS J0632+057, respectively.

 LS\,5039, discovered by \cite{1998A&A...338L..71M}, is
 a binary at a distance 
 $d=2.9\pm 0.8$ kpc 
 composed of a massive ($M_\ast=23$~M$_\odot$)
 O6.5V((f)) type star ($V=11.2$ mag) and a compact object 
with a poorly constrained mass, $\sim (1$--$4) M_\odot$. 
The compact object orbits the star with a period $P_{\rm orb}\simeq 
3.9$
days, which is the shortest orbital period among all HGMBs.    The binary
orbit inclination angle is $i\sim 30^\circ$ \citep{2005MNRAS.364..899C,Sarty11}. Radio observations have shown a persistent (over many binary periods) AU-scale
asymmetric extension around LS~5039 whose morphology varies with orbital phase \citep{2012A&A...548A.103M}.
  Initially the extension  was interpreted as   jets  from an accreting compact object, 
which led to a ``microquasar'' classification 
 \citep{2000Sci...288.2340P}.
However, more recently 
\cite{2012A&A...548A.103M} attributed the varying extended radio morphology to a pulsar wind nebula whose shape varies due to the interaction with the wind of the massive star. Overall, the debate over
whether the compact object is an accreting BH or a pulsar interacting
with its surroundings and producing an extended nebula
resembling ``jets'' is still ongoing (see e.g., \citealt{2015CRPhy..16..661D}). 

Obviously, the most direct evidence 
 of LS 5039 containing a pulsar would be the 
 detection of pulsations.  
 However, among all HMGBs radio
pulsations have only been detected in LS\,2883 and MT91\,213 \citep{2009ApJ...705....1C,2014MNRAS.437.3255S}. 
Even if LS\,5039 hosts a pulsar, the
non-detection of radio pulsations 
is not surprising 
because of the tight orbit and correspondingly large
optical depth to free-free absorption \citep{2006A&A...456..801D}. Therefore, searching for pulsations in X-rays may be a more promising approach for LS\,5039 than for the other HMGBs. 

Pulsations have been searched for in previous X-ray observations of LS 5039. For instance, \cite{2011MNRAS.416.1514R} found no periodicity in the {\sl CXO} observation of LS\,5039. 
However, \cite{Yoneda20}, 
 hereafter Y+20, have recently reported the detection of 
a period $P\approx 8.96$ s in the {\sl Suzaku} HXD data 
from 2007, and a potential counterpart at $P\approx9.05$ s in the \nustar\ data from 2016 (the same data we analyze below), at photon energies above 10 keV. 
The difference in the periods implies a fast spin-down,
suggesting that the compact object in LS\,5039 could be a magnetar (i.e., a NS with a very high magnetic field; \citealt{2017ARA&A..55..261K}). The statistical significance of this result is, however, questionable, as discussed below.

The shape of the hard X-ray spectrum can also provide critical
information about the nature of the compact object. For instance, some
NSs in accreting HMXBs show cyclotron resonant scattering features in
the range of about 10--100 keV 
and 
many of them have exponential cutoffs around a few tens of keV
(e.g., \citealt{2002ApJ...580..394C}). Such spectral characteristics are very
efficiently detected with {\sl NuSTAR} (see e.g., \citealt{2014ApJ...784L..40F,2016MNRAS.457..258T}). 
LS\,5039's
spectrum was  studied in the 0.01--10 MeV range with {\sl INTEGRAL},
{\sl RXTE}, {\sl Suzaku}, and {\sl CGRO}. 
Most of these measurements suggest a relatively soft
high-energy spectrum ($\Gamma\approx 2$), while the  {\sl Suzaku} HXD spectrum is more
consistent with a simple extrapolation of the 
$\Gamma=1.4$--1.6 PL spectrum
measured below 10 keV with 
{\sl CXO}, {\sl XMM-Newton}, and {\sl
  Suzaku} XIS
  \citep{2009ApJ...697..592T}

Of all known HMGBs, LS\,5039 has the shortest orbital period.
This prompted us to carry
out {\sl NuSTAR} observations of LS\,5039 over the  entire orbital period to obtain a
complete spectral and temporal portrait of this HMGB. Additionally, we used archival {\sl Suzaku} XIS observations 
that also cover the full binary period and extend  spectral coverage to lower energies.
In Section \ref{obs} we
describe the {\sl NuSTAR} and {\sl Suzaku} observations and data reduction procedures. In
Section \ref{timing_analysis} we describe the binary orbit corrections to the
arrival times caused by the compact object's motion around its massive
companion and present the results of the periodicity
 and  variability search.
  In Section \ref{specana} 
we present the spectrum of LS 5039 and the results of spectral fitting
as a function of the orbital phase. We discuss our findings and  conclude with a brief summary in Section \ref{summ}.

\section{Observations and Data Reduction}
\label{obs}

The {\it Nuclear Spectroscopic Telescope Array} (\nustar, \citealt{
  harrison13ApJ:NuSTAR}) consists of two 
 similar modules, FPMA and
FPMB, operating in the energy range 3--79~keV. The \nustar\ 
 observation  of  LS 5039 (ObsID 30201034002) started on 2016 September 1  
  (MJD 57632.0972)
and lasted for $T_{\rm obs} \approx 345\,{\rm ks}\,\,
\approx 1.024 P_{\rm orb}$ ($\approx60$ consecutive {\sl NuSTAR} orbits).
 We processed the data using the
\nustar\ Data Analysis Software, \texttt{nustardas} ver.\ 1.8.0. The photon arrival times were corrected to the solar system barycenter using the \texttt{barycorr} tool\footnote{See \url{https://heasarc.gsfc.nasa.gov/ftools/caldb/help/barycorr.html}.} and the latest clock correction file\footnote{nuCclock20100101v110.fits.gz \url{http://nustarsoc.caltech.edu/NuSTAR\_Public/NuSTAROperationSite/clockfile.php}}. 
The timing accuracy of {\sl NuSTAR} is expected to be 65 $\mu$s, on average \citep{2021ApJ...908..184B}. 
We
reduced the data using the \texttt{nuproducts} task 
and HEASOFT ver.\ 6.22.1. 

For spectral analysis and binary lightcurves we used the flags \texttt{-saacalc=2
  --saamode=optimized --tentacle=yes} to correct for enhanced
background activity visible at the edges of the good time intervals (GTIs)
immediately before entering the SAA. This resulted in a total GTI of about 166~ks.
We extracted source events 
from the 60\arcsec radius circle
 around the
source position,
 which
maximized the S/N ratio. Background events are extracted from an 
annulus around the source position with the inner and outer radii of
120\arcsec\ and 200\arcsec.

To extend 
the spectral analysis to 
  lower 
   energies, 
 we used 
  archival {\sl Suzaku} data.
  {\sl Suzaku} observed LS\,5039 between 2007 September 9  and 2007 September 15, with a total scientific exposure time of $\approx 203$ ks (obsID 402015010). The observation, originally reported 
  by \cite{2009ApJ...697..592T}, provided coverage of about 1.5 orbits of the  LS\,5039 binary.     In the soft X-ray energy band 
  (0.3--12 keV), {\sl Suzaku} had four X-ray telescopes \citep{2007PASJ...59S...9S} each with 
  its own focal plane CCD camera (X-ray Imaging Spectrometer; XIS; \citealt{2007PASJ...59S..23K}) having an $18'\times18'$ field-of-view. The XIS2 camera was turned off in November 2006 due to an anomaly and is not used in our analysis. The XIS0 and XIS3 detectors use front-illuminated CCDs, while the XIS1 has a back-illuminated CCD.   
  
  We used  HEASOFT ver.\ 6.25 
  for {\sl Suzaku} data reduction. The data were reprocessed using the {\tt aepipeline} script and were reduced following the standard procedures\footnote{See \url{https://heasarc.gsfc.nasa.gov/docs/suzaku/analysis/abc/}.}. The source spectra and light curves were extracted from a $3'$ radius circle centered on the source position, while the background spectra and light curves were extracted from a $3'$ circle placed $7\farcm5$ south of the source in 
  each of the three XIS images. The response matrix and ancillary response file 
  were made using the {\tt xisrmfgen} and {\tt xissimarfgen} tools, respectively. 
  Since there are  known calibration issues near the Si edge in the XIS detectors near 2 keV  (see e.g., \citealt{2011PASJ...63S.991S,2013ApJ...772...83L}),
   we exclude the 1.7--2.3 keV energy range from all of our spectral fits.
    Prior to producing the light curves, the event arrival times were corrected to the solar system barycenter using the {\tt barycorr} tool. Since the {\sl Suzaku} XIS time resolution was only 8 s in this observation, we did not use the XIS data for the periodicity search. 
  We 
  also used the (barycentered) {\sl Suzaku} HXD data 
  to investigate the candidate periodic signal  reported by Y+20, but we did not use them for spectral analysis
  (because of the strong background contamination in this non-imaging instrument).  All errors quoted throughout the paper are reported 
at the $1\sigma$ level, unless otherwise noted.

\section{Timing Analysis}
\label{timing_analysis}

A fully coherent periodicity search for an observation with a length comparable to the LS~5039 binary period is a ``needle in a haystack''  type problem.   The large uncertainties in the orbital ephemeris (see Table \ref{tab:orb}) require a 
prohibitively large 
grid in the multidimensional parameter space to guarantee that the 
periodic signal 
is not missed (see the discussion in  \citealt{2012MNRAS.427.2251C}). An alternative approach 
is to segment the observation into multiple time  intervals during  which the radial velocity of the compact object is approximately constant, and 
search for periodicity within each segment by analyzing the distribution of Fourier power in the time-frequency domain (a dynamic  power spectrum search; e.g., \citealt{2012hpa..book.....L}). 

Such an approach 
 was employed by Y+20 (who found a period candidate $P=9.05$ s in the \nustar\ data). 
These authors only searched for 
 a signal with 
 period $P>1$~s. They 
 justify this restriction due to the relatively small number of photons
 in the 10--30 keV band chosen for the periodicity search in the \nustar\ data. Furthermore, they  
 justify this energy band selection 
 by the fact that the 8.96 s period  candidate was seen in {\sl Suzaku} HXD, which 
 is only sensitive above 10 keV.
 However, 
 we see no reason for the signal not to be present below 10 keV 
 because there is no change in the source spectrum 
 (see Section \ref{specana}). 
 Since we 
 cannot exclude the possibility
 that the true period is 
 different from that claimed by Y+20, we 
 perform a period search in a broader range of frequencies and a 
 different 
 energy range.
 
To improve the sensitivity by decreasing the spread of the potential signal in the frequency domain, we (1) use an optimal division into time segments 
that depends on  the frequency intervals we are searching in, (2) introduce a statistic that provides a higher sensitivity to a signal than the simplistic incoherent summing of Fourier powers from non-overlapping time intervals within the observation,   and (3) apply the R\"omer delay correction (e.g., \citealt{1976ApJ...205..580B}) to the photon arrival times using the best known ephemeris. Then we perform the dynamic pulsation search. 
This approach allows us to search up to much higher frequencies 
(e.g, 1\,000 Hz) 
than the 1 Hz limit used in Y+20.

Since the nature of the compact object is unknown,  
we adopt the approach, 
described in Section \ref{dynamicfourier}, which allows us to search for both 
periodic signals and 
quasi-periodic oscillations (QPOs). We also perform a burst-like variability search (on scales from 1 s to 200 s)
and orbital variability characterization on larger timescales (Section \ref{bin_lc}). 
  
\begin{deluxetable*}{ccccccc}[t!]
\tablecaption{Orbital parameters for LS\,5039 inferred from different observations
\label{tab:orb}}
\tablewidth{0pt}
\tablehead{
& \colhead{Aragona09}  &  \colhead{Casares11} & \colhead{Sarty11} &
\colhead{Yoneda20} & \colhead{Yoneda20} &
\colhead{this work} \\
 &   &   &  &
\colhead{\sl NuSTAR} & \colhead{Suzaku} &
}
\startdata  
$P_{\rm orb}$ (d) & $3.90608\pm0.00010$ & $3.90608\pm0.00008$   & 3.906           &  3.90608 & 3.90608 & $3.9057\pm0.28$ \\
$T_0$ (MJD)       &	$52825.48\pm0.05$	& $52477.58\pm0.06$     & $55016.58\pm0.06$ 
& $57629.250_{-0.030}^{+0.023}$   & $54352.455_{-0.035}^{+0.05}$ & 
$57633.71\pm0.06$ \\ 
$e$               & $0.337\pm0.036$     & $0.35\pm0.03$         & $0.24\pm0.08$ 	
& $0.306_{-0.013}^{+0.015}$ & $0.278_{-0.023}^{+0.014}$  & $0.289\pm0.09$       \\
$W_\ast$ (deg)       & $236\pm5.8$         & $212\pm5$             & $237.3\pm21.8$    
& $236.8_{-3.1}^{+2.3}$  & $234.6_{-3.3}^{+5.1}$ 	&			\\ 
$a_\ast\sin i$ (lt-s) & $3.33\pm0.15$   & $4.06\pm0.16$         & $4.11\pm0.35$	&&&\\	
$W_p$ (deg)     	&  	$56\pm5.8$        & $32\pm5$	        & $57.3\pm21.8$	 
& $56.8_{-3.1}^{+2.3}$ & $54.6_{-3.3}^{+5.1}$ & $44.6\pm4.1$ \\
$a_p\sin i$ (lt-s)  &  	$48\pm14$      & $52_{-19}^{+9}$	    & $52_{-19}^{+10}$	
& $48.1\pm0.4$ & $53.05_{-0.55}^{+0.7}$ & $48.1\pm2.7$   		
\enddata
\tablecomments{Orbital parameters from \cite{2009ApJ...698..514A}, \cite{Casares11}, \cite{Sarty11}, 
 Y+20 and this work. Subscripts $\ast$ and $p$ correspond to the massive star and the compact object (putative pulsar), respectively. For the \cite{2009ApJ...698..514A}, \cite{Casares11},  and \cite{Sarty11} orbital solutions, the projected semi-major axis 
$a_p\sin i$ 
was calculated assuming 
 $m_\ast = (23\pm 3) M_\odot$
while
the compact object's mass
was assumed to be $m_p =(1.6\pm0.4) M_\odot$ for \cite{2009ApJ...698..514A} and $m_p=1.8_{-0.6}^{+0.2}M_\odot$  
for \cite{Casares11} and \cite{Sarty11}. We use the \cite{Casares11} orbital parameters 
for the R\"omer delay correction. The Y+20 
and `this work' parameters were not measured from observations of the massive star but 
obtained from fits maximizing the significance of the putative period near 9.05 s. 
The large uncertainties of the `this work' parameters take into account the multitude of solutions of about the same significance in the vicinity of the best-fit solution (see Section \ref{p9s_cand}).
}

\end{deluxetable*}

\subsection{Correction for the R\"omer delay caused by the orbital motion}

As the putative pulsar
is orbiting a massive star, the distance between the pulsar and the observer changes with  orbital phase, which translates into changing times of photon travel 
to the observer. 
This effect can be equivalently described in the observer's frame as a 
Doppler shift of the pulsation frequency 
varying with the binary phase because of
the changing radial velocity of the
pulsar.
Different authors have inferred slightly different sets of orbital parameters from optical observations of the massive companion (see examples in Table \ref{tab:orb}).
To correct the event arrival times 
for this effect (R\"omer delay), we adopted the
`eccentric fit + 1d oscillation' orbital solution from
 \cite{Casares11},
which included 
modulation of the radial velocity 
with a 1 day period, possibly caused by non-radial oscillations of the massive companion in its eccentric orbit.
The orbital dependencies of the R\"omer delay and Doppler shift 
of the other published binary solutions are within $\approx \pm 1\sigma$ uncertainties of the \cite{Casares11} curves for these quantities (see Figure \ref{fig:deltat} 
and Figure \ref{fig:vel_acc} in the Appendix). 

The projection of the orbit's semi-major axis 
onto the sky plane, $a\sin i$, and the longitude of periastron, $W$, in 
\cite{Casares11} (and the other papers quoted in Table \ref{tab:orb})
pertain to the massive stellar component of the binary.
 They are connected with the corresponding compact object 
 parameters as follows, 
\begin{equation}
	a_p\sin i=
	\frac{m_\ast}{m_p}\, a_\ast\sin i\,,
	\quad\quad
	W_p=W_\ast-180^\circ,
\label{eq1}
\end{equation}
where the subscripts $\ast$
and $p$ correspond to the massive
star and the compact object (putative pulsar), respectively, $m$ is mass, and  $i$ is the orbital inclination (the angle between the orbital plane and the plane of the sky). 
For the pulsar (NS)  mass we use $m_p=1.8_{-0.6}^{+0.2}M_\odot$,  from the assumption that $m_p$ should be in the range of 1.2--2.0 solar masses, which gives $a_p\sin i=52_{-19}^{+9}$\,lt-s = $(1.55_{-0.6}^{+0.27})\times 10^{12}$\,cm.

The corrections to the photon arrival times  due to orbital motion (R\"omer delay)
are calculated as follows,
\begin{eqnarray}
	t_\text{corr}=t-\frac{a_p\sin i}{c}\left[\sin W_p(\cos E-e)+\right.\nonumber\\\left.
	\sqrt{1-e^2}\cos W_p\sin E\right]\,.
\label{eq:romercorr}
\end{eqnarray}
Here $e$ is the eccentricity of the orbit, and $E$ is the eccentric anomaly,
\begin{equation}
	E-e\sin E=\Omega_{\rm orb}(t-T_0)\,,
	\label{exc_anomaly}
\end{equation}
were $\Omega_{\rm orb}=2\pi/P_\text{orb}$, 
$P_\text{orb}$ is the orbital period, 
and $T_0$ is the epoch of  periastron \citep{1976ApJ...205..580B}. The right hand side of Equation (\ref{exc_anomaly}) is commonly called the mean anomaly.
 Depending on the orbital phase, the correction ranges from 
$-$40 s to $+$60 s (see Figure \ref{fig:deltat}).

\begin{figure}[t]
\centering
  \includegraphics[width=0.5\textwidth]{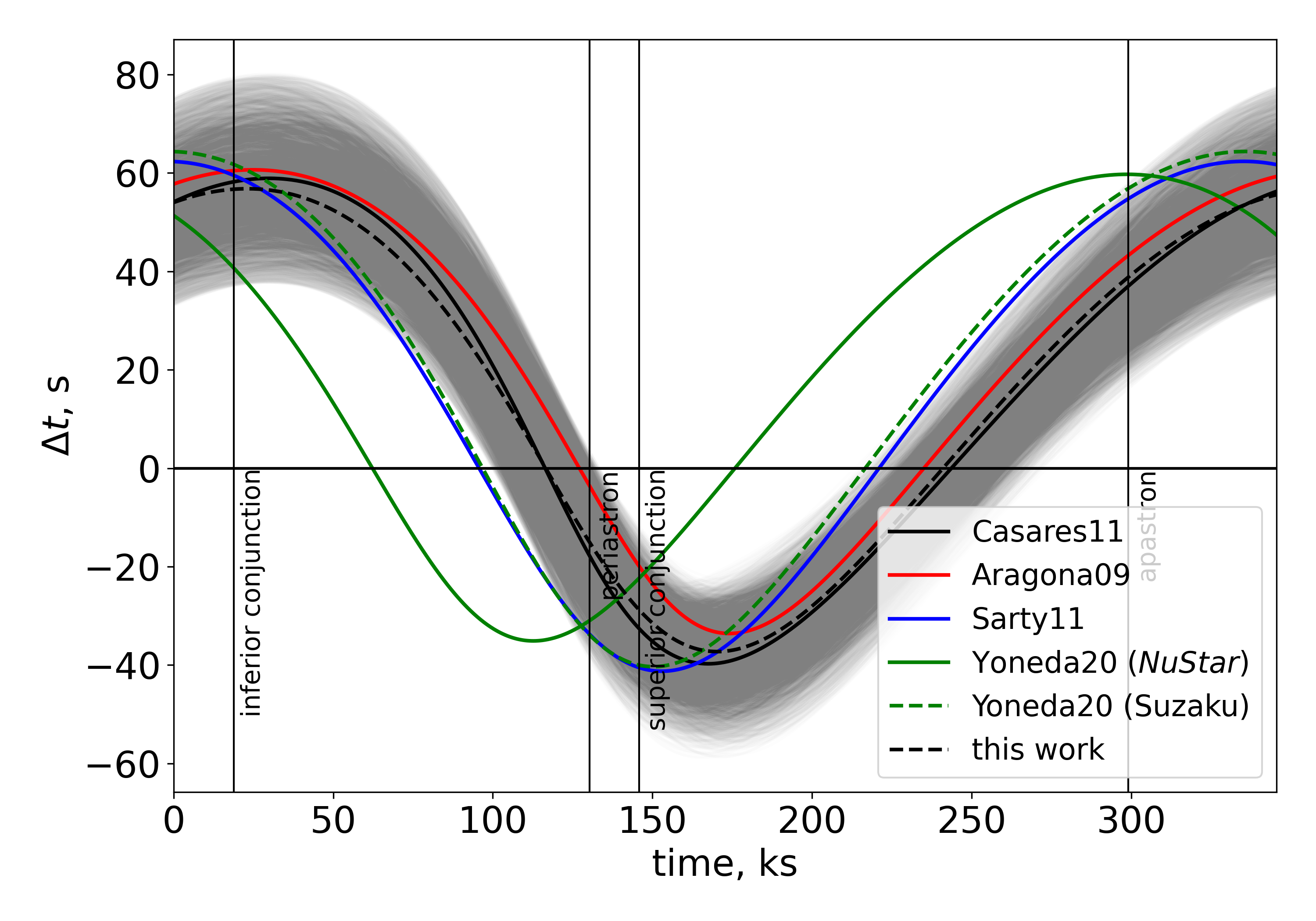}
  \caption{Arrival time corrections due to R\"omer delay during our {\sl NuSTAR} observation for 6 
  sets of orbital parameters (see Table \ref{tab:orb}). Shown is $\Delta t = t_\text{corr}-t$ vs.\ $t$, given by Equation (\ref{eq:romercorr}). The shaded area shows the  uncertainty of $\Delta t$ due to the  uncertainties of the orbital parameters for the binary ephemeris from \cite{Casares11} (the model with 1\,d oscillations).
 }
  \label{fig:deltat}
\end{figure}

\subsection{Periodicity search in the \nustar\ data}
\label{sec:perser}

We searched for periodic or quasi-periodic signals up to $f_{\rm max}=1000$ Hz
by analyzing   the arrival times of $N=56\,647$ events, 
registered during $\approx190$ ks (when the target was not occulted by the Earth),
in the photon energy range of 3--20 keV.
We excluded higher energies to reduce the background contamination as \nustar's sensitivity decreases at higher energies. 
We corrected the event
arrival times 
for the R\"omer delay using the best-fit binary parameters 
from Casares et al.\ 2011
(Table \ref{tab:orb}) prior to the search. 

\subsubsection{
Fourier power spectrum for the entire \nustar\ observation 
assuming the binary parameters are known with high precision
}
\label{fourier_entire_obs}
\begin{figure*}[]
\includegraphics[width=1\textwidth]{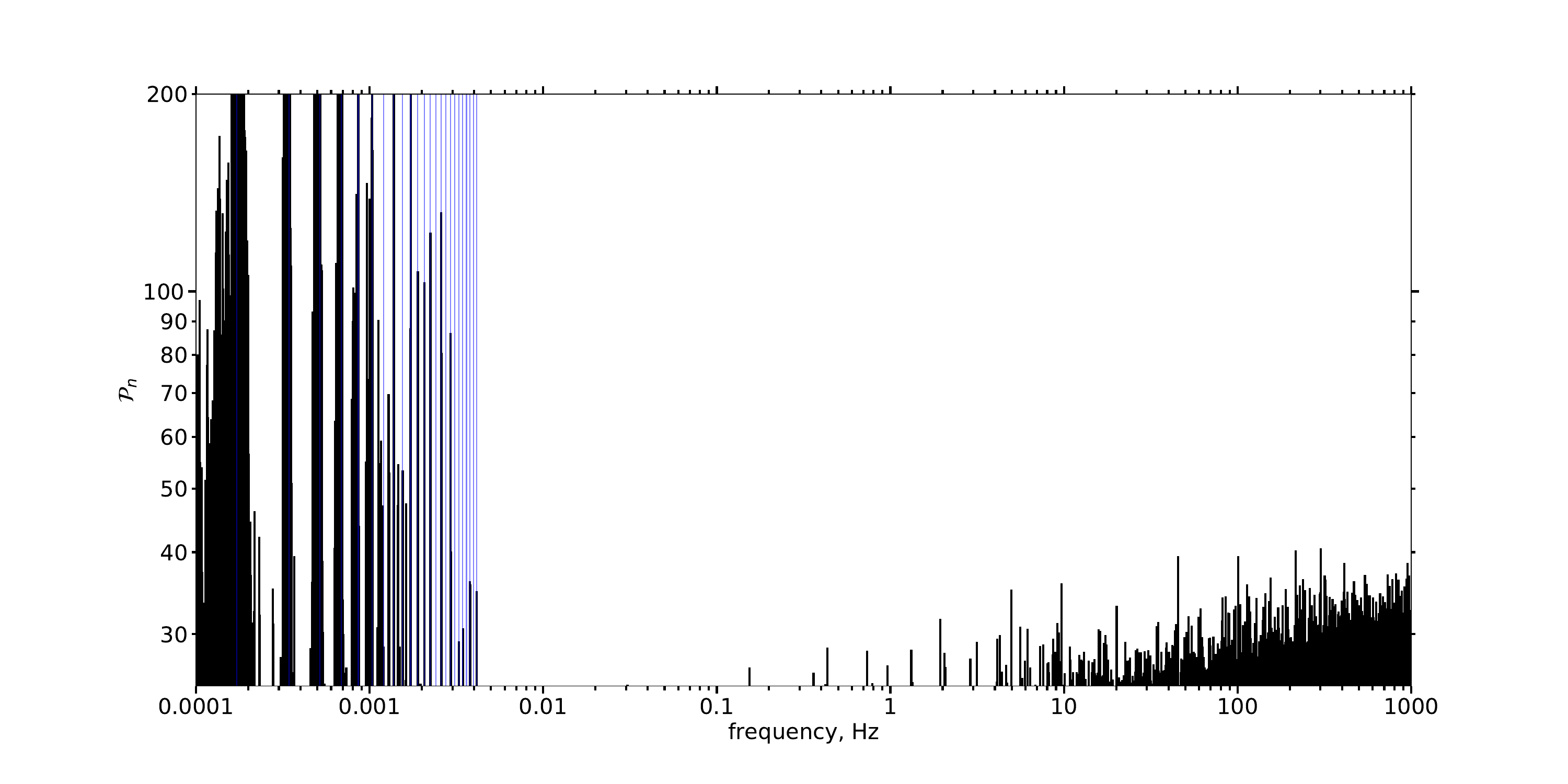}
\includegraphics[width=1\textwidth]{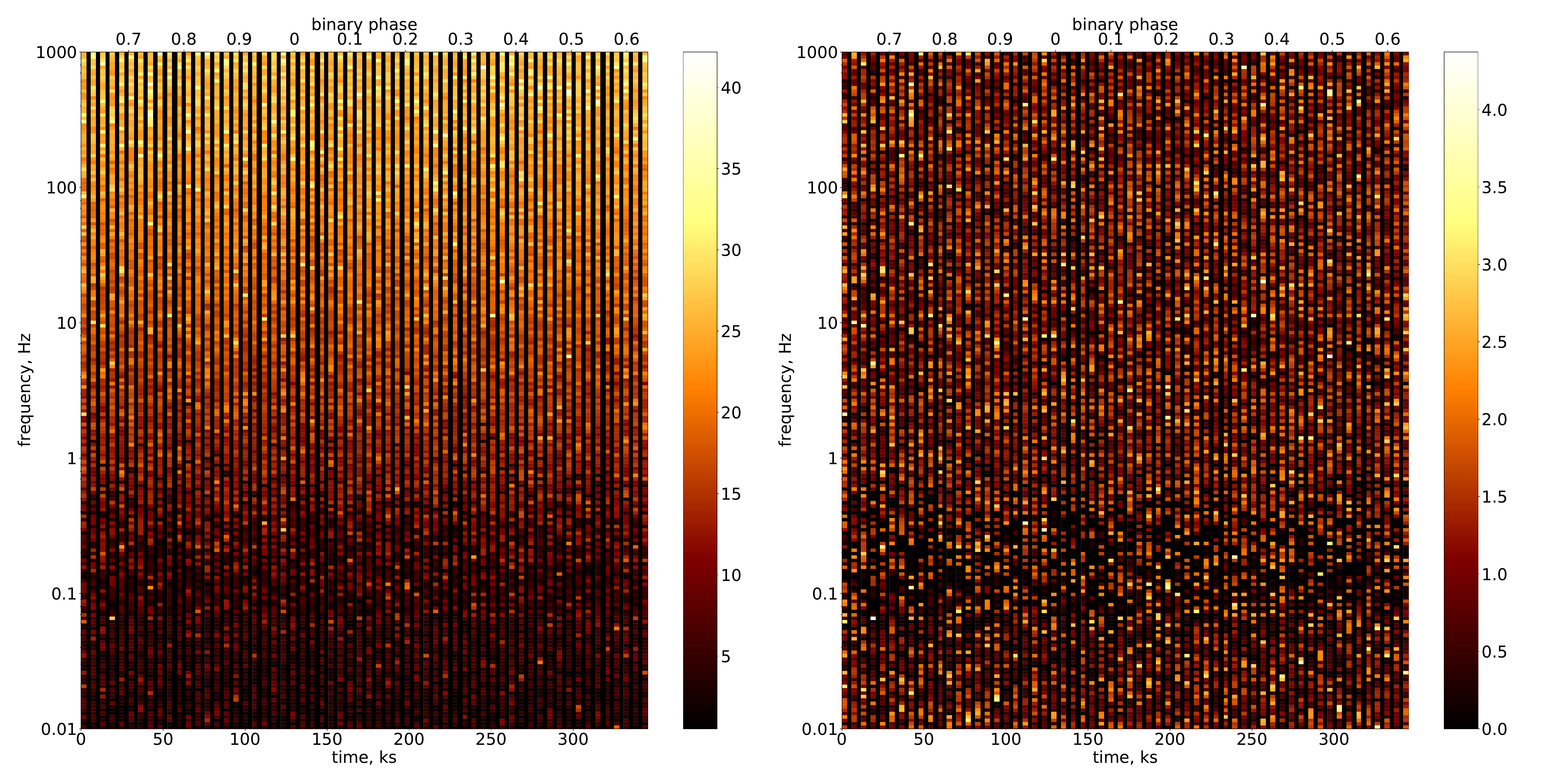}
  \caption{
  The top panel shows
   the Fourier power spectrum ${\cal P}_n$, 
   calculated from the entire \nustar\ observation
    (see 
    Appendix \ref{appendixA} and Section \ref{fourier_entire_obs}).
    Prior to calculating the Fourier power spectrum, the  arrival times of photons (with energies restricted to the 3--20 keV band) were corrected for the R\"omer delay 
using the best-fit orbital parameters from \cite{Casares11} (see    Table \ref{tab:orb}).     The vertical blue lines correspond to the \nustar\ orbital period and its harmonics. 
    The bottom panels show the 
     dynamic (time-resolved) spectra of Fourier power
     and significance,
      calculated individually for each  of the 60  \nustar\  orbits, after applying the same R\"omer delay correction (see Section \ref{fourier_spec_sep_orbs} and Appendix \ref{appendixA}).
      To improve the visualization quality,
      we do not show 
      power (and significance) values in each frequency bin
      but 
      instead divide the entire log-frequency range in 200 segments of equal size 
      and plot (using  color) the maximum Fourier power, ${\cal P}_k^{\rm max}$ (left), within the $k$-th segment and its significance, $\alpha_k$ 
      (right).
   The vertical black stripes are the gaps due to the occultation of LS 5039 by the Earth. 
    }
\label{fig:fur}
\end{figure*}

Using the corrected arrival times from the entire \nustar\ observation, we 
calculated the Fourier power 
spectrum 
(see the top
panel of 
Figure \ref{fig:fur} and the Appendix \ref{appendixA}).
The top
panel of Figure \ref{fig:fur} shows the values of the Fourier power ${\cal P}_n$ as a function of frequency $f=n\,\Delta f$, where $\Delta f \simeq 3\times 10^{-7}$ Hz is the 
frequency bin width 
(only 24,448 power values with ${\cal P}_n>20$ are shown).
The power is normalized in such a way that the mean $\overline{{\cal P}_n} = 2$ for a Poisson-distributed noise.
The 
high values of ${\cal P}_n$ 
at low frequencies, $f\lesssim 0.003$ Hz, are associated with the periodic motion of the \nustar\ satellite around the Earth (the frequency of \nustar\ revolution, $1.72\times 10^{-4}$ Hz, and its 
15 harmonics are shown by 
blue vertical lines). 
At higher frequencies 
none of 
the ${\cal P}_n$ values is  outstanding, and their significances do not exceed $4\sigma$.
Thus, we conclude that
no significant periodic signal is detected. 

To look for quasi-periodic signals, we increased the width of frequency bins by factors of 10, 100, and 1000 (compared to the natural width
$T_{\rm obs}^{-1}\approx 2.89$ $\mu$Hz)
but found no outstanding peaks in the binned 
Fourier power spectra.  

\subsubsection{
Fourier spectra in separate \nustar\ orbits }
\label{fourier_spec_sep_orbs}

The apparent lack of pulsations in the Fourier power spectrum of the entire observation could be due to 
a large difference between the actual  binary parameters and 
the parameters used for the R\"omer delay correction.
Because of this difference, the binary-phase-dependent frequency shift may not be fully compensated by the R\"omer delay correction. As a result, the signal coherence  would be lost, the power peak corresponding to the (unknown) pulsation frequency  would be spread over many frequency bins, and the peak's height  would be strongly reduced. 
To mitigate the coherence loss, 
we 
search for pulsations in much shorter time segments, corresponding to the intervals of visibility of LS\,5030 in the 60 consecutive  \nustar\ orbits covered by our observation.
During the relatively short intervals (3.2 ks on average)  
the difference between the actual and best-fit radial velocities  of the compact companion 
does not change as  much as over the entire orbit, and
there is a higher chance that the signal coherence is preserved. 

Similar to the search in the entire \nustar\ observation, we calculated 60 Fourier power spectra 
and 
  corresponding signal detection significances.
The results
%of this dynamical timing 
are shown in the bottom
panels of Figure \ref{fig:fur} as  time-frequency images in which the  power and detection significance values are shown by the 
brightness of the image ``pixels''.
The vertical light and dark stripes correspond to intervals of visibility and occultation of LS\,5039, respectively.

If pulsations were detected, they would be seen as a sequence of brighter (lighter) pixels along the time axis at frequencies around the pulsation frequency. If the actual binary parameters coincided with the 
assumed ones, this sequence would be seen as a horizontal stripe parallel to the time axis. Deflections of the brighter pixels from a horizontal stripe would provide the difference between the 
 actual values of the radial velocity and the ones used in the R\"omer delay correction, in separate \nustar\ orbits. 

We see from the bottom-right panel of Figure \ref{fig:fur} that the search in separate \nustar\ orbits also did not provide periodicity detection. 
Although some of the $\alpha_k$ values may appear marginally significant, one has to keep in mind that $\alpha_k$ is defined for a single \nustar\ orbit (see the Appendix \ref{appendixA}), and  the true  significance is therefore lower when all orbits are included (due to the larger number of statistical trials). Moreover, we do not see any extended 
(along the horizontal direction) connected clusters of adjacent pixels.   

The lack of detection could be due to the lower sensitivity of this search. The first reason for the sensitivity loss is the  smaller number of counts in separate orbits 
than in the entire observation (on average, 994 versus 56,647 counts). Since, for a periodic signal with a given pulsed fraction, the power ${\cal P}_n$ is proportional to the number of counts, 
weak pulsations would not be detected.

The second reason for the sensitivity loss is the spread of signal frequency over several frequency bins caused by the Doppler shift.
In the $i$-th \nustar\ orbit, the 
spread associated with the Doppler shift
unaccounted  for by the R\"omer delay correction can be estimated as $(\delta f)_i \sim f\, |\Delta \dot{v}_{\parallel,i}| T_i/c$, where $T_i$ is the visibility interval, and $\Delta \dot{v}_{\parallel,i}$ is the difference between the 
assumed and actual radial accelerations of the binary motion in the middle of the $i$-th orbit (see Appendix \ref{appendixB}). This spread becomes greater than the natural width $T_i^{-1}$ of the frequency bin for $f > \tilde{f}_i \sim c\, (|\Delta \dot{v}_{\parallel,i}| T_i^2)^{-1} = 
3.9\, [|\Delta \dot{v}_{\parallel,i}|/(3\,{\rm m\,s}^{-2})]^{-1} [T_i/(3.15\,
{\rm ks})]^{-2}$ Hz.
At a frequency $f$ substantially higher than 
$\tilde{f}_i$ the  peak in the signal power is spread over $f/\tilde{f}_i$ bins, and the peak height will be reduced by about the same amount.

\subsubsection{Search for pulsations in a dynamic Fourier  spectrum with frequency-dependent time windows}
\label{dynamicfourier}
Since the R\"omer delay correction is 
 imperfect
due to the binary ephemeris uncertainties,
a periodic signal 
 with a certain frequency $f_0$ in the reference frame of the pulsar
 can be spread 
 by the Doppler effect
 over 
 a number of neighboring frequency bins in the observer's reference frame
(see  the Appendix 
\ref{appendixB} for details).
This spread can be mitigated by 
splitting the observation duration $T_{\rm obs}$ into $N_w$ shorter time windows of a length $T_w = T_{\rm obs}/N_w$ (hence wider frequency bins $T_w^{-1}$), but then the signal 
can be shifted to different frequencies $f$ in different time windows.
  In order to maximize the sensitivity to such a signal, one should  
select optimal lengths of the time windows 
  and use an efficient algorithm for detecting the signal 
  in the time-frequency domain.

The optimal lengths 
of the time 
windows are determined by the
requirement that the maximum possible drift in frequency does not exceed $T_w^{-1}$.  Such  choice of
$T_w$ is optimal because if one chooses an even smaller $T_w$, then  events  are  lost  (as the  number  of  events  is  proportional to $T_w$) and the Fourier power decreases. 

As we show in Appendix \ref{appendixB}, the optimal lengths and numbers of time windows depend on the signal frequency
($T_w\sim 9.9 f_0^{-1/2}$ ks, $N_w\sim 35 f_0^{1/2}$ in our case). 
For practical purposes,
it is convenient to divide the entire 0--1000 Hz frequency range into 7 broad frequency intervals with
different numbers and lengths of time windows.
For each of these time windows we calculate the Fourier power spectrum ${\cal P}_n$ in the 0--1000 Hz frequency range, with a frequency resolution 
of $T_w^{-1}$.

As the next step,
we split the entire frequency range 
into 
segments 
   $ f_m(1-\beta) < f < f_m(1+\beta)$, 
where the coefficient $\beta$ is proportional to $\delta v_\parallel/c$,
$\delta v_\parallel$ is the maximum residual uncertainty of the pulsar's radial velocity in the appropriate time window, and the
central frequency $f_m$ of the $m$-th segment
($m=1, 2, \ldots$) satisfies Equation (B13). 
The segment width $2\beta f_m$ must be large 
enough to ensure that the entire signal,
  whose Doppler-shifted frequency $f$ is 
  varying with time, 
 is contained within this 
 segment (see Figure \ref{fig:mc2} in the Appendix).  

Each of these segments (their total number is about 2880, 
for the chosen $\beta=0.0022$
in the 0.01--1000 Hz range)
is inspected  for the presence of signal signatures. 
Any chosen segment
is within one of the seven broad frequency intervals, described in the Appendix \ref{appendixB}, which determines 
the number $N_w$ and length $T_w$ of the time windows
(hence the choice of the precalculated Fourier power spectra) appropriate for the segment analysis.

In the time-frequency plane, an $m$-th frequency segment
consists of $N_w$ time windows and $N_f = 2\beta f_m T_w$ frequency bins, i.e., of $N_w\times N_f= 2\beta f_m T_{\rm obs} = 690 (\beta/10^{-3}) f_m$ elements
for which Fourier powers 
${\cal P}_{n,j}$ have been calculated ($n$ and $j$ number the frequency bins and time windows, respectively). 
To locate 
possible signal signatures in the segment, we pinpoint the largest values ${\cal P}_{j}^{\rm max}$ 
in the sets of $N_f$ powers ${\cal P}_{n,j}$ within each of the time windows $j$. 
If the ${\cal P}_j^{\rm max}$ values represent a sufficiently strong signal, then the mean of these values over the entire
time of observation
\begin{equation}
\mu=\frac{1}{N_w}\sum_{j=1}^{N_w} {\cal P}_{j}^{\rm max},
\label{eq:mu}
\end{equation}
should significantly exceed a similarly defined mean for the noise,
$\mu_{\rm noise}$.

To characterize the significance of a possible excess of the measured $\mu$ over $\mu_{\rm noise}$,
   we introduce the following $s$-statistic: 
\begin{equation}
	s=(\mu-\mu_\text{noise})/\sigma_\text{noise}\,,
	\label{eq:stat}
\end{equation}
which  
provides the signal significance in units of standard deviation.  Here, $\sigma_\text{noise}$ is the standard deviation of the noise.

Because the observed data contain numerous time gaps, and the count rate 
changes with time,
we used Monte-Carlo simulations to infer $\mu_\text{noise}$ and  $\sigma_\text{noise}$  for each  $N_w$ 
(when simulating noise,  we included gaps larger than 60 s).

\begin{figure*}[hbt]
  \includegraphics[width=0.48\textwidth]{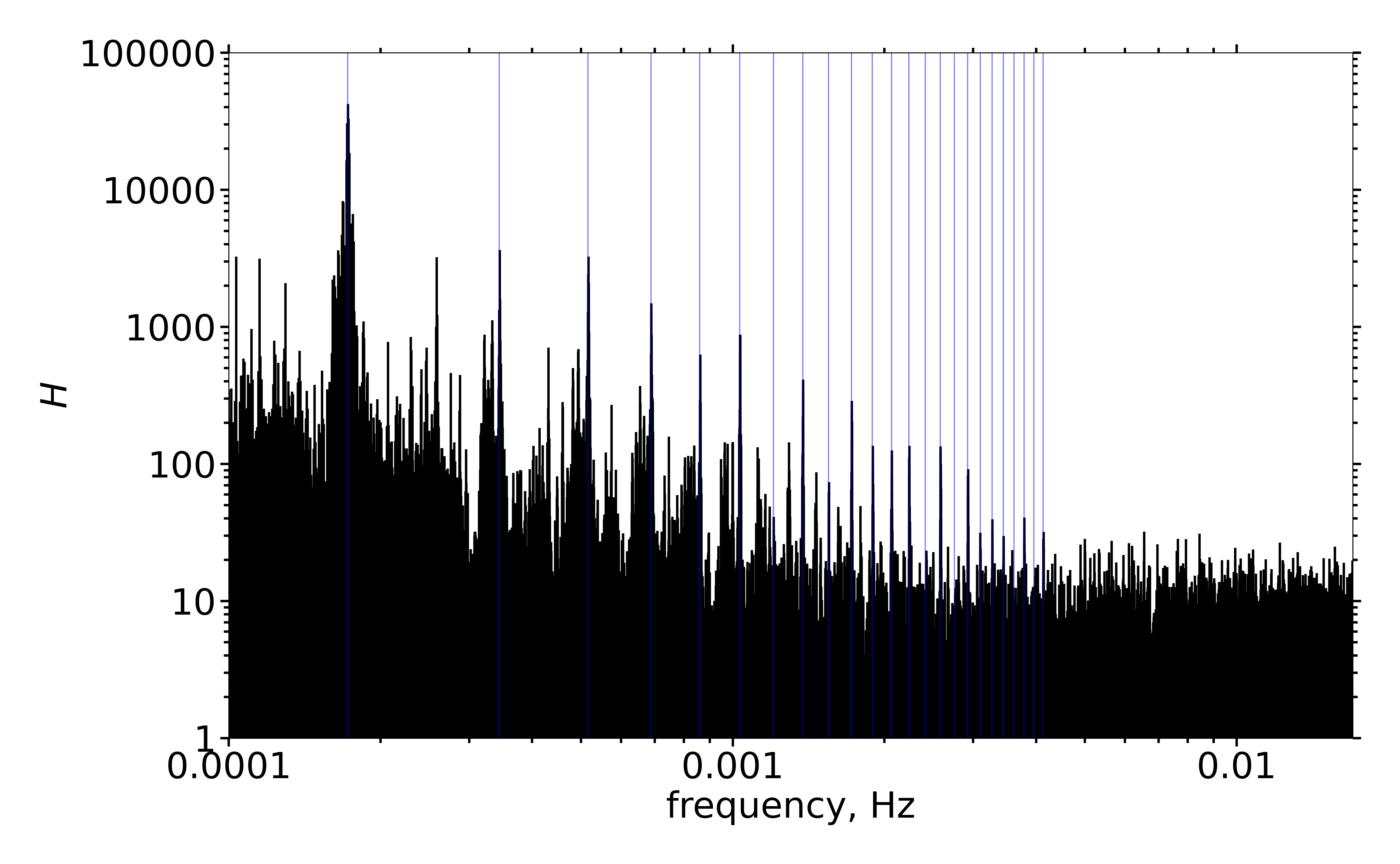}
  \includegraphics[width=0.48\textwidth]{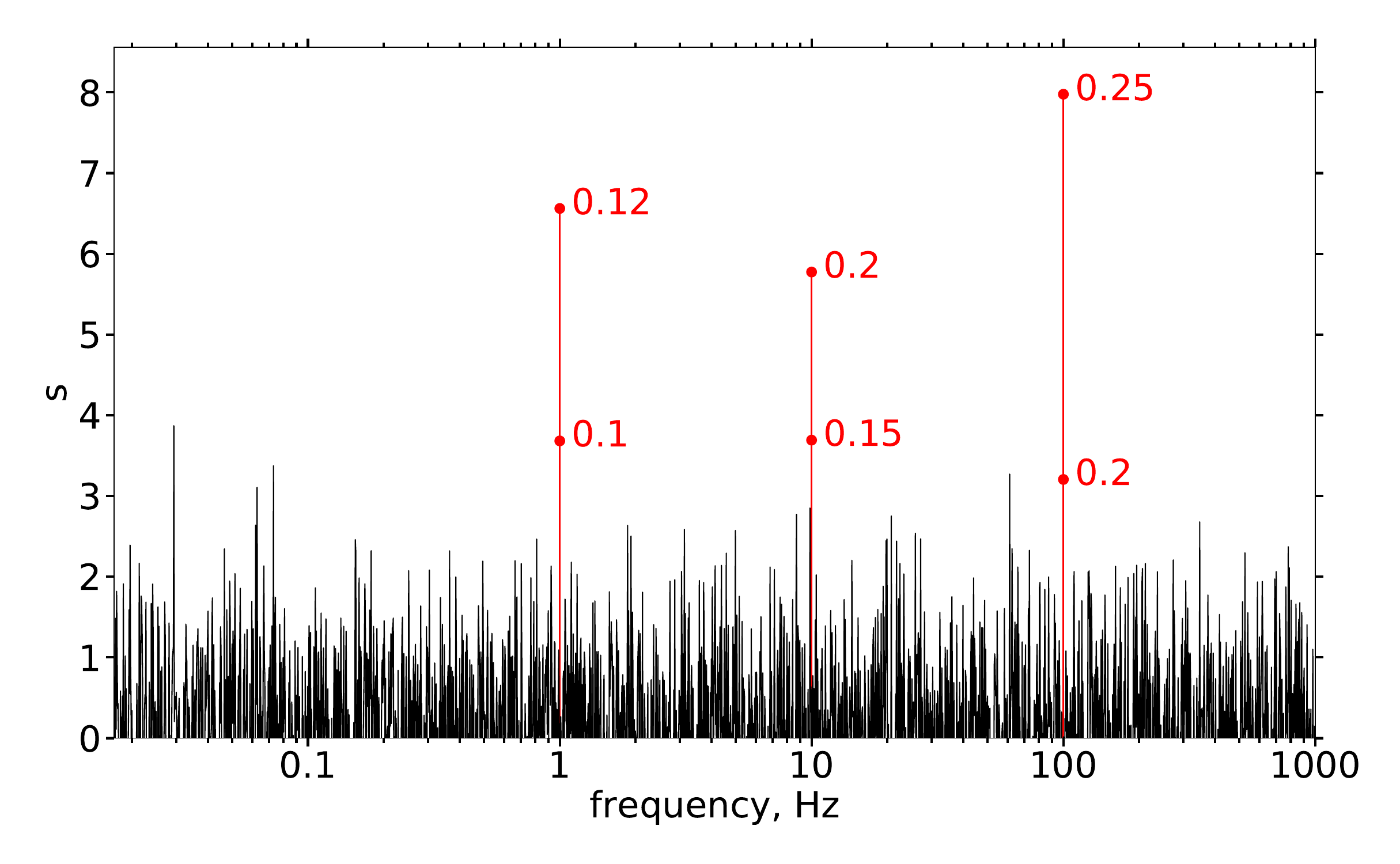}
  \caption{
   Left panel shows the $H$ statistic 
   calculated for the frequencies up to 0.017 Hz. 
   Blue vertical lines  correspond to the orbital frequency of \nustar\ 
($1.7\times 10^{-4}$ Hz)
and its harmonics. 
The H-statistic peaks are aliases caused by the presence of quasi-periodic gaps (the gap width 
varies slightly with time) due to the occultation of the target by Earth during the long \nustar\ observation
 and SAA passages. 
Right panel shows all positive $s$-statistic values  (see Equation \ref{eq:stat}) in 1261  frequency segments
between 0.017 Hz and 1000 Hz.
Dots on red vertical lines show the $s$ values for the (artificial) test signals (see the description and Figure (\ref{fig:mc2}) in the Appendix \ref{appendixB}); the numbers near the dots are the signal fractions.
}
\label{fig:stat}
\end{figure*}

In the right panel of Figure \ref{fig:stat} 
 we plot
the $s$-statistic values for
  for each
  of 1261 frequency segments in which $s$ is positive
(the total number of frequency segments is 2496 
in
the 0.017--1000 Hz range). 
  Although there are several peaks  slightly exceeding the $3\sigma$ level (i.e., $s>3$),
 no signal candidates are passing the $4\sigma$ threshold.

  This panel also shows the $s$-statistic values computed for six simulated periodic
  sinusoidal test  signals with varying levels of   
  signal strength (characterized by the 
 signal fraction $p$, with $1-p$ being the   unpulsed
  fraction)
  that are imperfectly corrected for the R\"omer delay
  (see the time-frequency images in Figure \ref{fig:mc2} in Appendix \ref{appendixB}). 
We see that the detectability of a signal with a given signal 
fraction strongly depends on  the signal's frequency (the higher the frequency the larger $p$ must be for the signal to be detected). At plausible young pulsar frequencies $\sim 3$--100 Hz, the signal fractions 
 should significantly exceed 0.1--0.2 to be detected with this method in the available \nustar\ data. 
   
   For the low-frequency part of the Fourier power spectrum,
    (Figure \ref{fig:stat}, left panel) we 
   use the more conventional $H$-statistic,
   defined as $H=\max\{Z_m^2-4m+4\}$ for $1 \le m \le 20$ \citep{1989A&A...221..180D}, where $Z_m^2$ is the statistic commonly used in periodicity searches in X-ray and $\gamma$-ray astronomy 
   (see, e.g., Buccheri et al.\ 1983). 
This is possible  because for $f\lesssim 0.01$ Hz the residual drift due to the imperfect R\"omer delay correction 
   does not  exceed the natural width ($\sim T_w^{-1}$) of the peak in $H$-statistic spectrum 
   during the entire observation (i.e., $N_w=1$; see the Appendix \ref{appendixB}). Although the $H$-statistic values at lower frequencies are high, they mostly coincide with multiple integers of the \nustar\ orbital frequency (shown by blue vertical lines) while the others are likely to be  aliases due the visibility gaps that are somewhat varying in their duration. No credible signal is detected.     
     
\subsection{$P=9.05$ s candidate} 
\label{p9s_cand}

Y+20
searched the {\sl Suzaku} HXD data of 2007 September 
 for 
 periodic 
   signals with 
 periods $P>1$~s in the 10--30 keV band. 
 They found 
 maximum $Z_{4}^2=68.0$ (the other $Z_n^2$ values were not reported)
 at $f_{\rm HXD}=0.1116510(5)$ Hz, or $P_{\rm HXD}=8.95648(4)$ s, with the estimated
 significance of 98.8\%.

In a subsequent  search in the 
above-described {\sl NuSTAR} data 
(191 ks net observing time, 12,000 events in the 10--30 keV energy band, and assuming $P>1$~s) 
Y+20 obtained a maximum $Z_{4}^2$=66.9 at
$f_{\rm NuST} = 0.1104507(4)$ Hz or $P_{\rm NuST}=9.05381(3)$ s,  with an  estimated significance of only 93\%.
 
 To find these period values, Y+20 varied the orbital parameters in the ranges provided in Table 1 of that work, applied the R\"omer delay correction for each parameter set, and chose the set, and the corresponding period, that maximized the $Z_4^2$ value.
The uncertainties of the periods given in Y+20  appear to be underestimated because they do not account for the uncertainties in the ephemeris parameters (as we show below).
 Y+20 concluded that two  
 incompatible  sets of  orbital parameters are needed to maximize the  strength of the {\sl NuSTAR}  and {\sl Suzaku} candidate periodic signals.
 Moreover, the ephemeris 
 that maximizes the periodic candidate signal in the {\sl NuSTAR} data
 is incompatible with  any previously published ephemeris (within their uncertainties;  see Table \ref{tab:orb} and Figure \ref{fig:deltat}).

 \begin{figure}[t]
 \centering
  \includegraphics[width=0.55\textwidth]{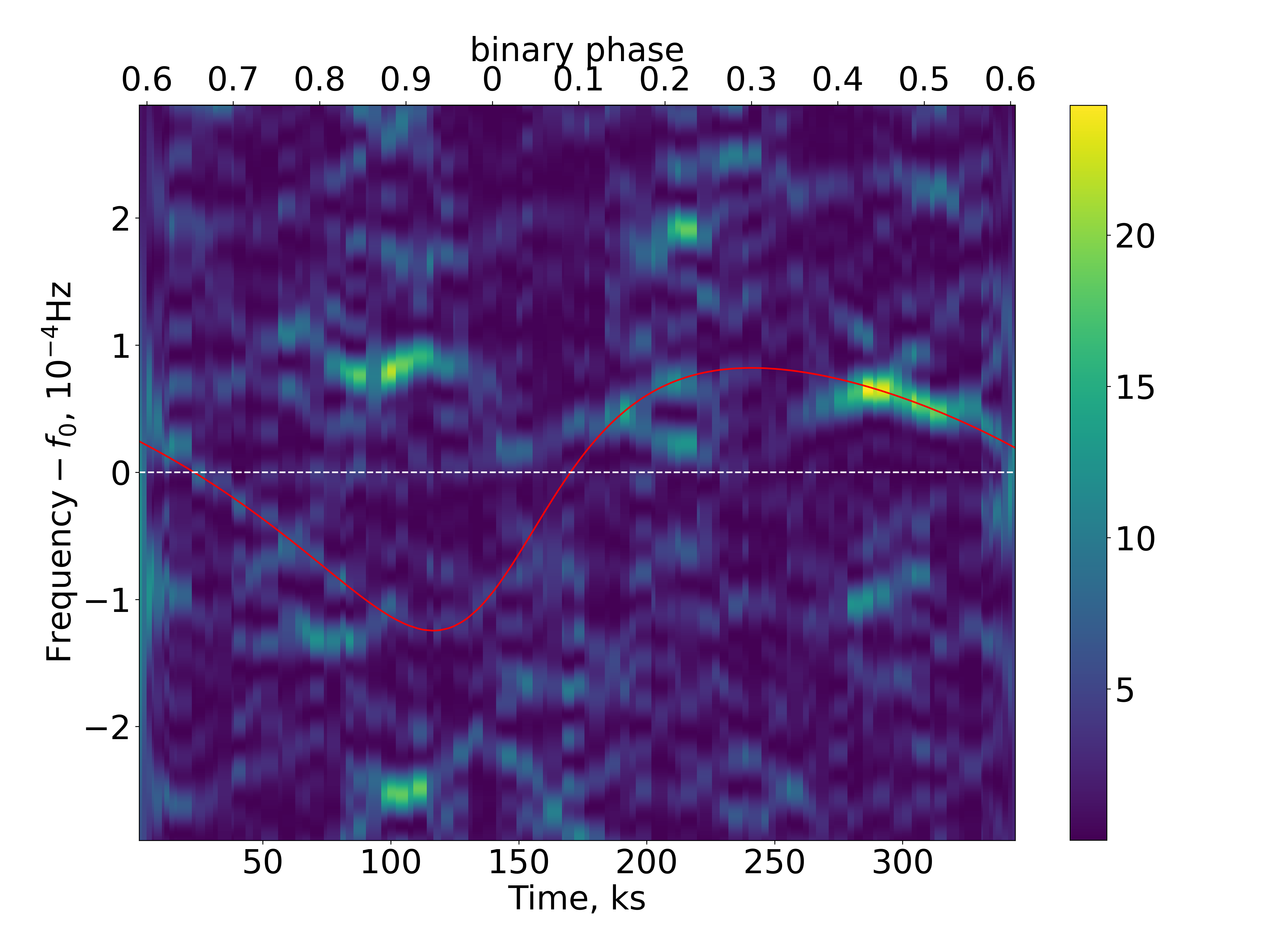}
  \includegraphics[width=0.55\textwidth]{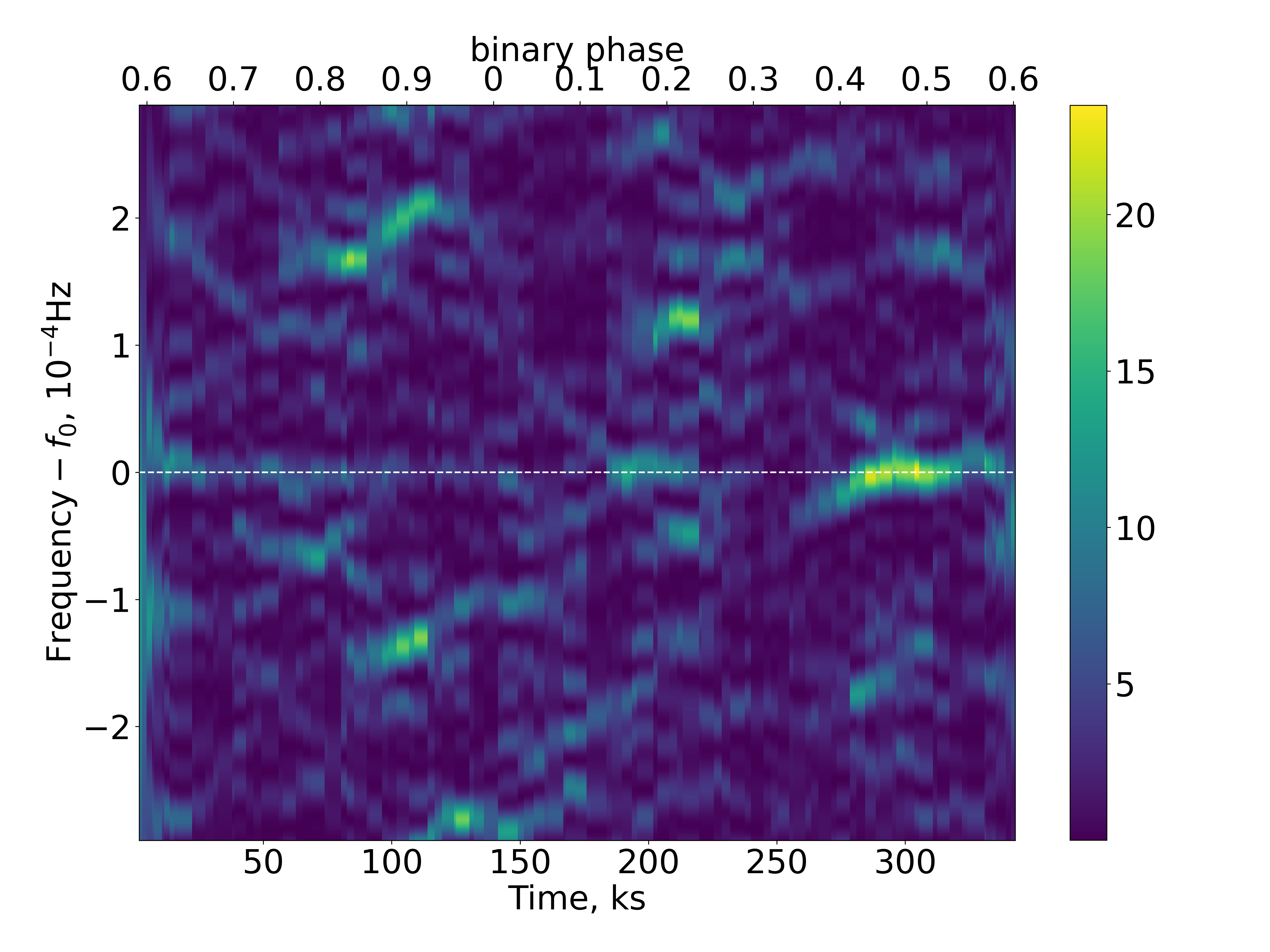}
  \caption{ 
  The top panel shows $Z_1^2$  as a function of 
  frequency and 
   time near
  $f_0=0.110498$ Hz ($P_0=9.04996$ s),  shown by the dashed horizontal line. 
  The $Z_1^2$ was calculated using events extracted 
  from the $r=38''$ aperture in the 10--18 keV band.
   The red curve shows the 
  Doppler shift due to the orbital motion as a function of the observation time 
  for the 
  orbital parameters that we found (see Table \ref{tab:orb}).
    The bottom panel shows the same thing as the top panel but the photon  arrival times  are corrected for the orbital motion with this ephemeris. 
 In both cases the color bars show the value of $Z_1^2$ 
 per time window.
 }
  \label{9ssignal_1}
\end{figure}

\begin{figure}[t]
 \centering
  \includegraphics[width=0.6\textwidth]{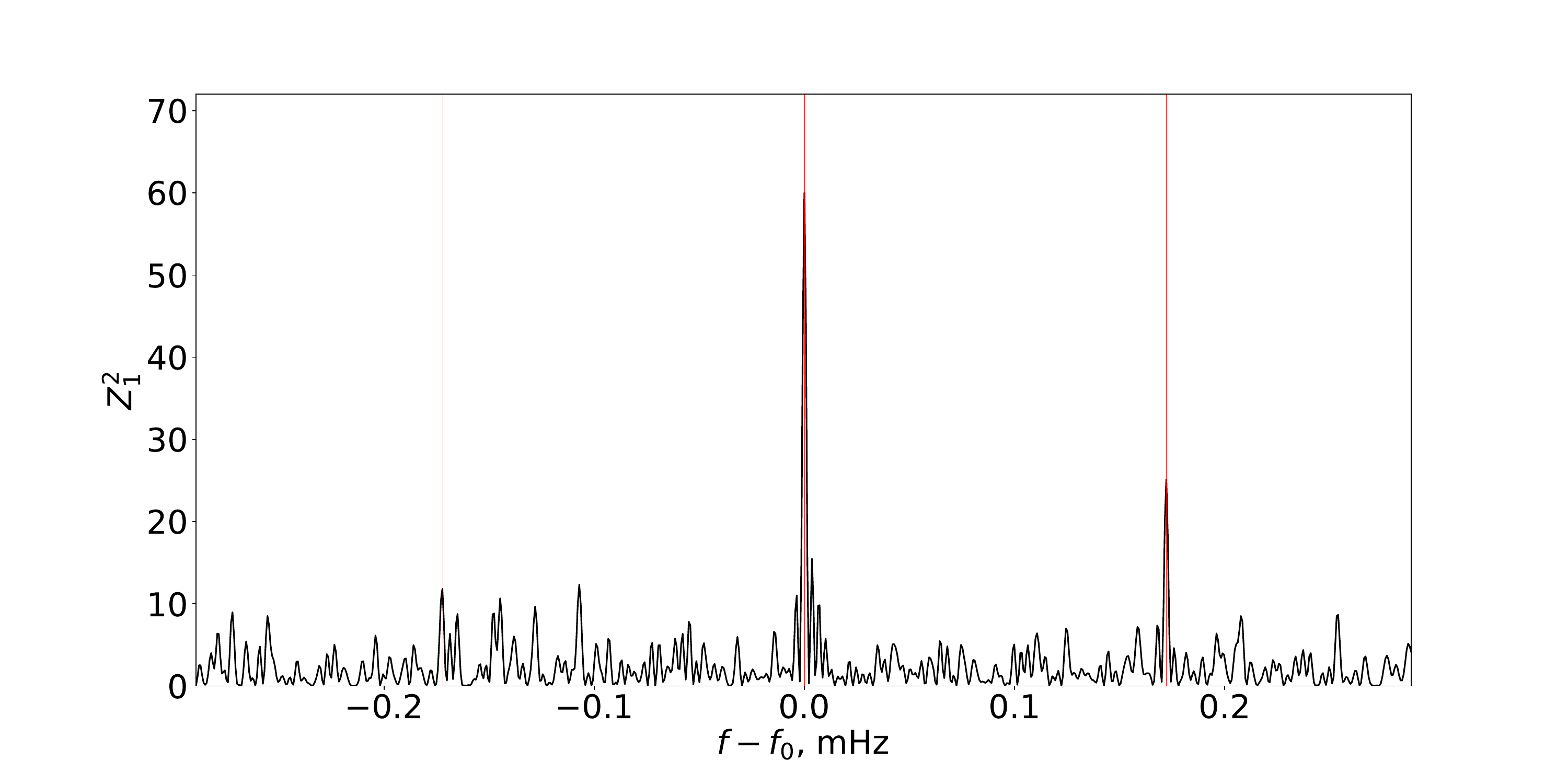}
  \includegraphics[width=0.55\textwidth]{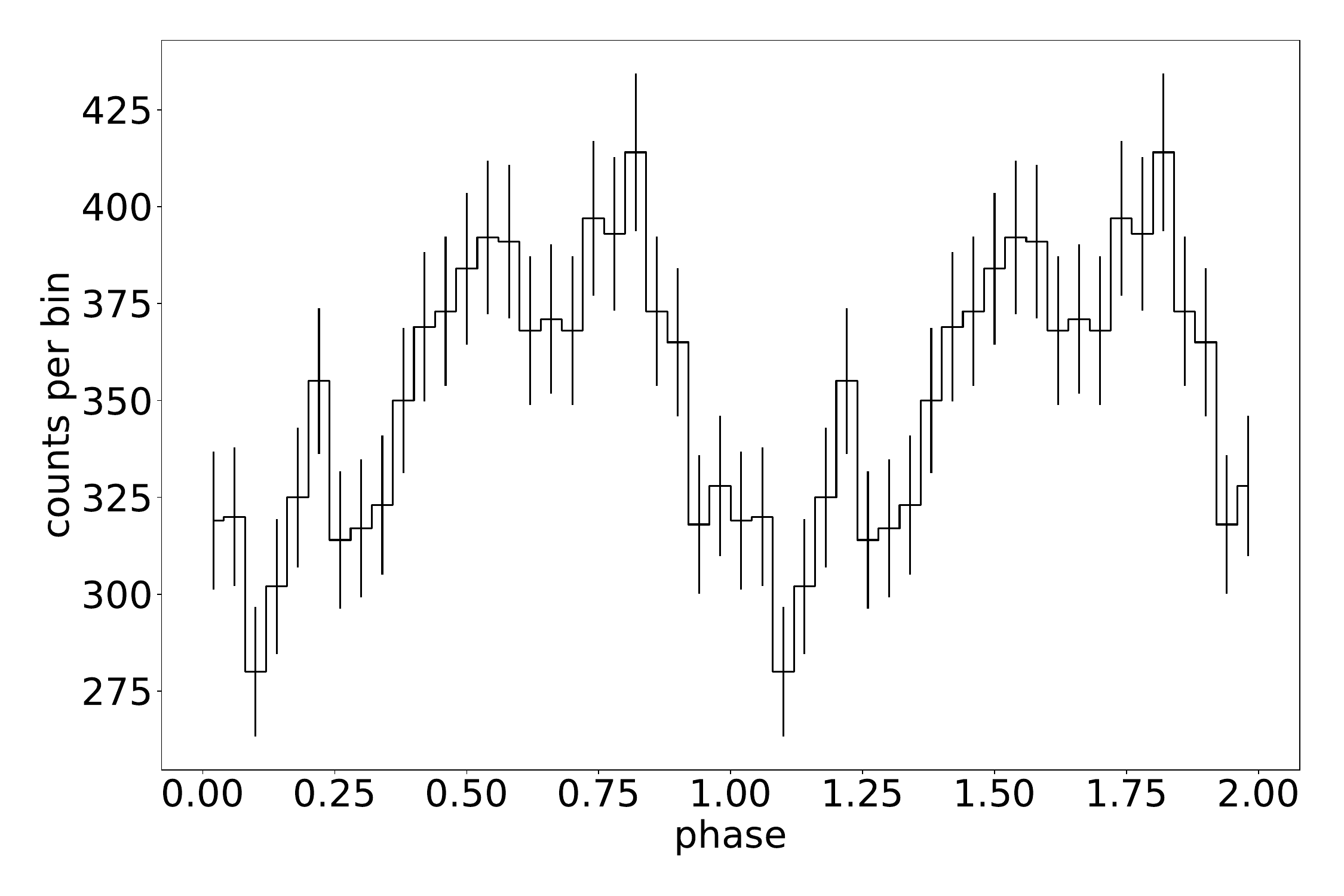}
  \includegraphics[width=0.55\textwidth]{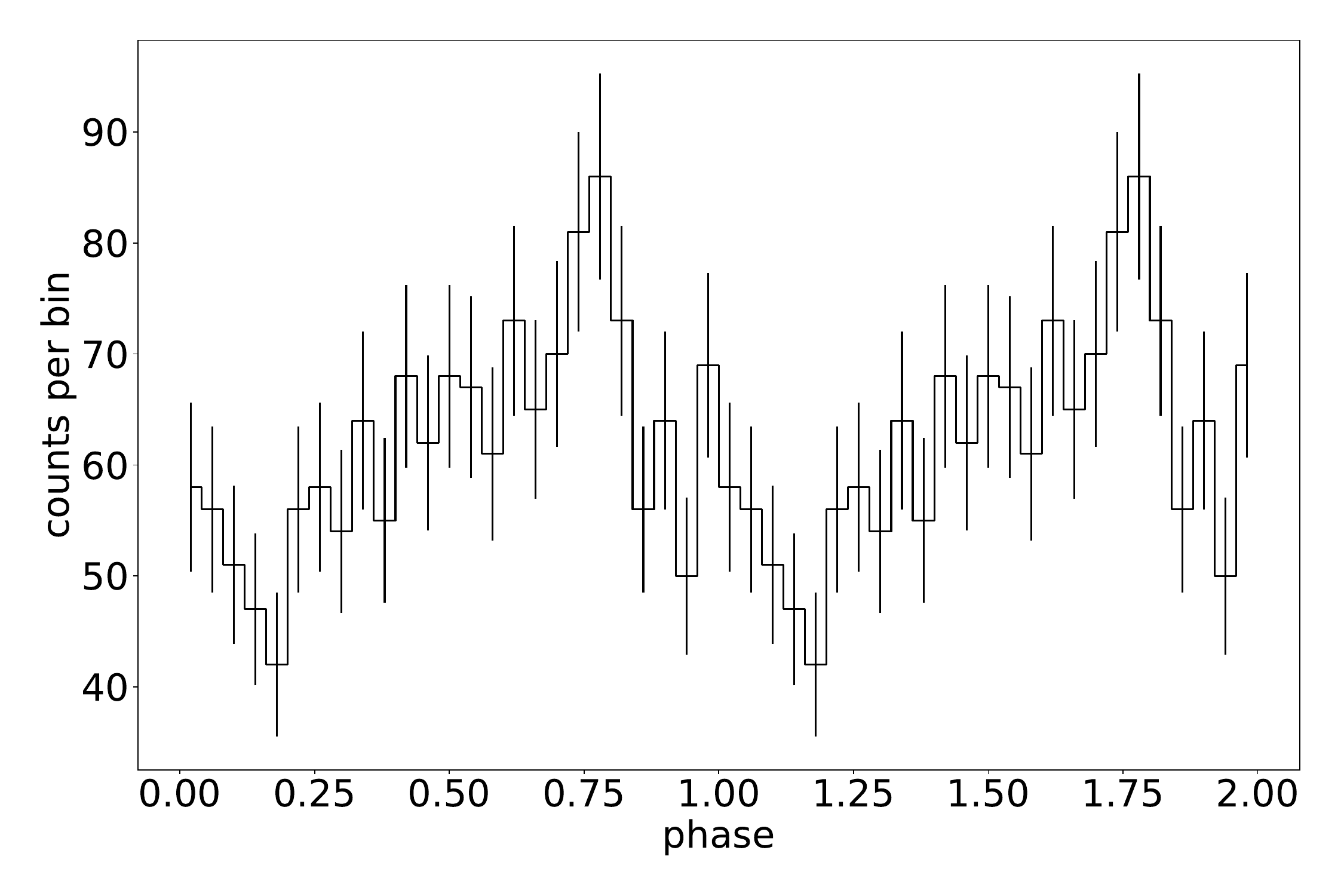}
  \caption{ The top panel shows $Z_1^2(f)$ 
  in the vicinity of  $f_0=0.110498$ Hz
  ($P_0=9.04996$ s) calculated 
  for the entire observation 
  with arrival times corrected for the 
  orbital motion with 
  the same ephemeris as in Figure \ref{9ssignal_1}. The vertical red lines mark the aliases of the main peak  which are offset from  $f_0$
  by $\pm 0.017$ mHz, which is the frequency of the \nustar\ orbit around the Earth.
  The middle and bottom panels show the pulse profiles 
  folded on the 
  period $P_0$ for the entire observation and for the 270--320 ks interval 
  (counting from the beginning of the observation), where the signal is particularly strong
  (see Figure \ref{9ssignal_1}). 
 }
  \label{9ssignal_2}
\end{figure}

To verify the period and the  significance  
reported by Y+20, we have performed 
 an independent period search 
in the {\sl NuSTAR} data  
around $P=9.05$ s. 
Choosing a set of 5 orbital parameters within the $2\sigma$ uncertainties of the Casares et al.\ (2011) ephemeris,  we applyed the R\"omer delay correction to the 
times of arrival,
calculated $Z_{m}^2(f)$ for $m=1,\ldots, 20$ (using 
events from the entire observation) in the frequency interval of (0.109--0.113) Hz, and used the $H$-test \citep{1989A&A...221..180D} 
to determine the maximum number of significant harmonics.
Varying the orbital parameters on the 5-dimensional grid, we found the parameter set that maximizes $Z_m^2$ and the corresponding frequency.
We repeated these calculations for various energy ranges within 3--40 keV and various source aperture radii up to $r=60''$ to choose an optimal (maximizing $Z_m^2$) energy range and aperture radius. 
As a result, in  the  10--18 keV energy range and for the aperture radius $r=38''$,
we found a  signal (shown in Figure \ref{9ssignal_2}) with maximum $Z_1^2=60$
at $f=0.1104977$ Hz ($P=9.049962$ s).
In the $Z_m^2(f)$ dependence, the peak at $f=0.1104977$ Hz  
is surrounded by many other peaks  with slightly lower heights, including a peak at 
$f=0.1104507$ Hz 
reported by Y+20. These 
peaks, appearing at 
different combinations of orbital parameters, look virtually as significant as the highest one, so that we cannot prefer one peak to another. Therefore, the true uncertainty of the putative pulsation frequency (and the fitted orbital parameters) is determined not by the width of a separate peak but by the width of the entire `cluster of peaks', $\sim 3\times 10^{-5}$ Hz in our case,  which is about two orders of magnitude larger than the uncertainty claimed by Y+20. 
Accounting for this uncertainty, the frequency and period of the putative pulsations are   
$f=0.11050(3)$ Hz and $P=9.050(2)$ s), for the orbital parameters listed in the column `this work' of Table \ref{tab:orb}. 

 We  note that higher harmonics are not required by the  $H$-test.
However,  for comparison with Y+20, who reports $Z_4^2$, we also calculated it and found  $Z_4^2=71$ for our best fit, at $f= 0.1104977$ Hz.
We also note that
 our optimal  
orbital parameters are much closer to those 
found in the previous papers
than the set of parameters 
suggested by Y+20
(see Figures \ref{fig:deltat} and \ref{fig:vel_acc}).

  To explore the distribution of the Fourier power as a function of 
  time (or binary phase), we calculated
   a time-resolved Fourier spectrum, 
 i.e.\  the $Z_1^2$
  distribution in the time-frequency plane around $f=0.110498$ Hz  
  ($P=9.04996$ s).
 To reduce the effect of time gaps, we used 
 a sliding time window with the size of $T_{\rm obs}/10$ which is moved by 10\% at each step 
  (see Figure \ref{9ssignal_1}). 
  The plots show that the strongest contribution to the signal comes 
  from a time interval close to the end of the \nustar\ observation (between 270 and 330 ks, counted from the start of the  observation), near the binary apastron.

We note that the 
$s$-statistic, introduced in Section \ref{dynamicfourier}, is not sensitive to such a signal because the Fourier power does not remain constant during the observation time span. 
Figure \ref{9ssignal_2} shows the $Z_1^2$ distribution calculated from the entire observation 
around $f=0.110498$ Hz ($P=9.04996$ s) and the corresponding folded pulse profiles from the entire observation and from the part of it with the strongest signal.

 In addition,  we re-analyzed the {\sl Suzaku} HXD data and confirmed   
 the signal candidate reported by Y+20, with maximum $Z_4^2 =67.8$ at $P=8.95648$ s.
 We also analyzed jointly the {\sl NuSTAR} and {\sl Suzaku} HXD data, requiring a common binary ephemeris. However, the strongest signal that we were able to find in this case was 
 rather insignificant, with  $Z_1^2\approx50$ 
 ($Z_m^2$ with $m>1$ were even less significant). 

It should also be noted that, among other factors (such as the energy range, aperture, and ephemeris choices),  the  significance of the $9.05$ s periodic signal candidate  depends  on the maximum frequency of the frequency range searched. 
There is no physical a priori reason to limit the search to 
  $f<1$ Hz. 
  Extending the frequency range   to much higher frequencies and accounting for 
  the huge number of trials associated with varying the ephemeris parameters, 
  energy range, and extraction aperture would  render the putative $9.05$ s  signal candidate  insignificant.
  
\subsection{
The binary light curves and search for nonperiodic variability}
\label{bin_lc}

\begin{figure}[t]
\begin{center}
  \includegraphics[angle=0,width=0.56\textwidth]{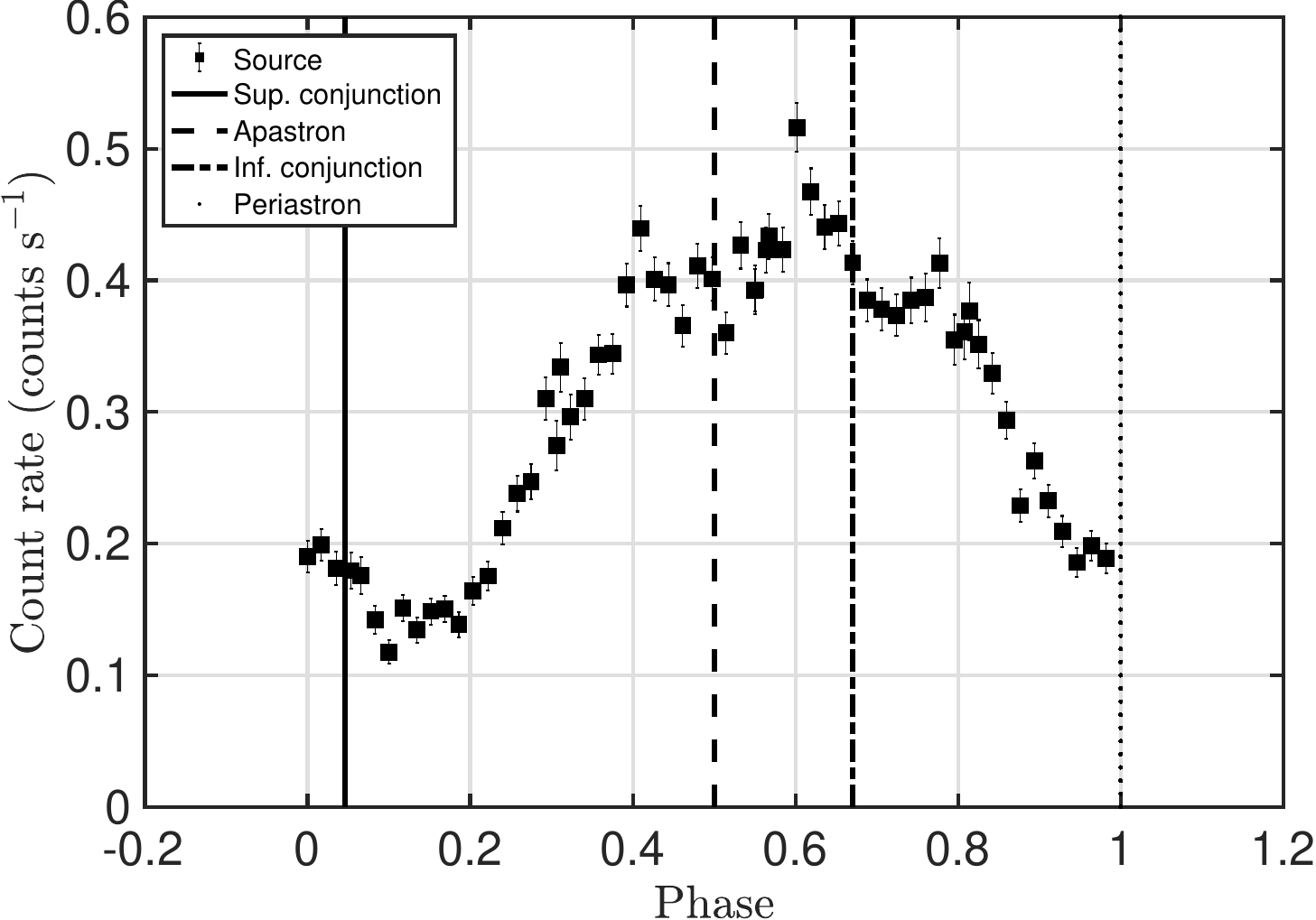}
  \includegraphics[angle=0,width=0.56\textwidth]{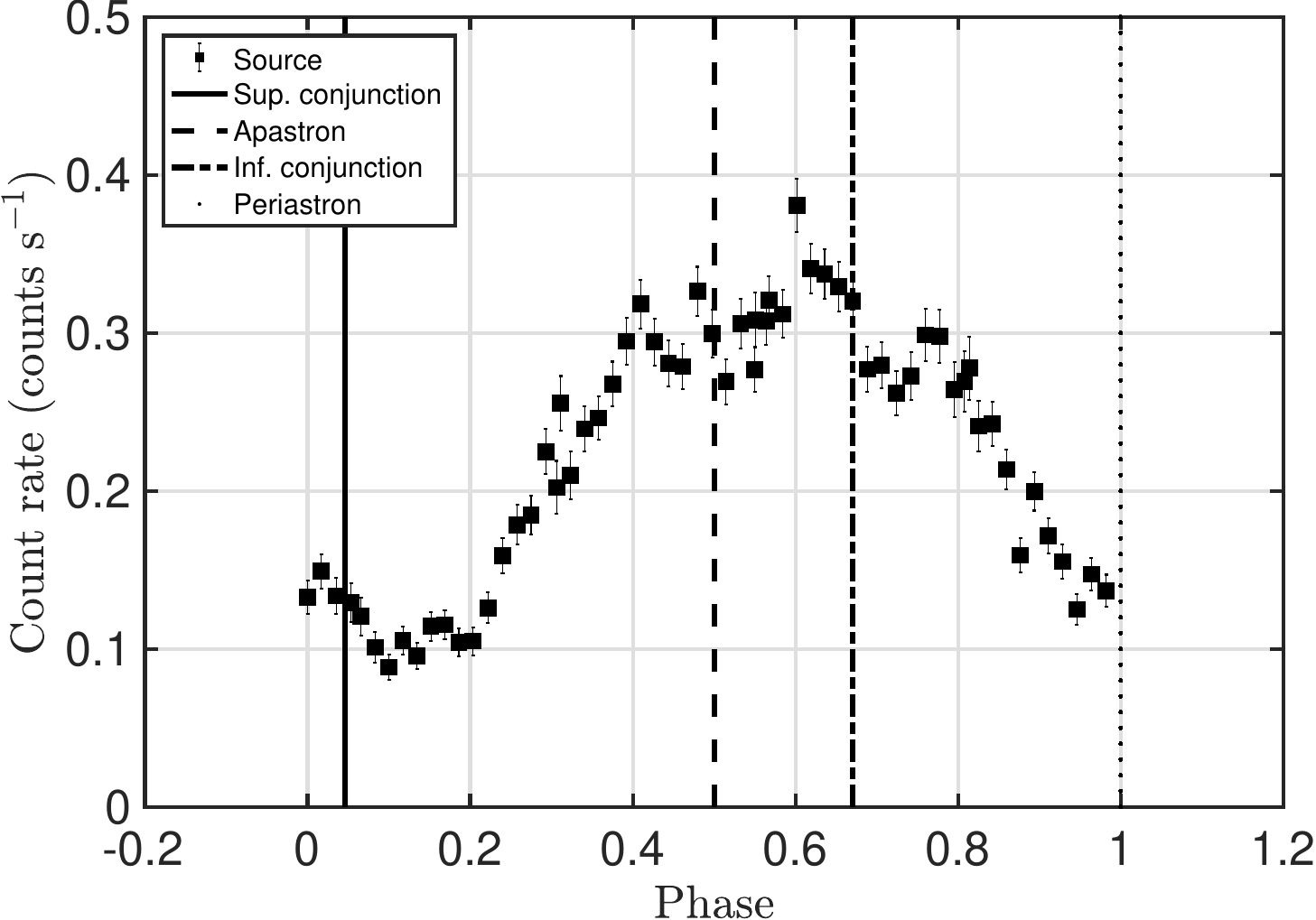}
  \includegraphics[angle=0,width=0.56\textwidth]{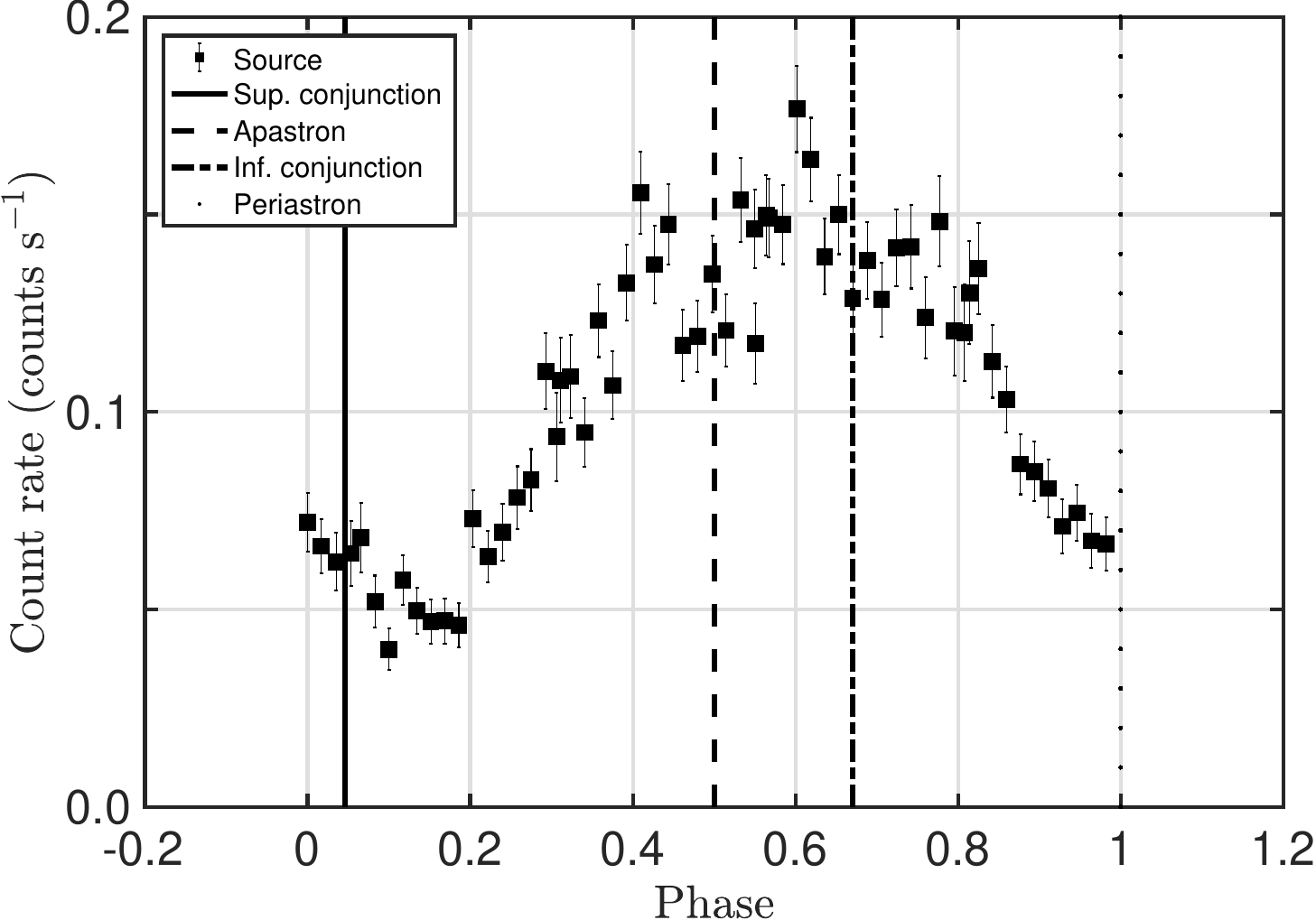}
\caption{\nustar\ background-corrected light curves 
of \src\ 
  as functions of orbital phase. 
The  upper, middle, and lower
    panels are for the energy
bands 3--60~keV, 3--10~keV, and
  10--60~keV. The superior and inferior conjunctions are 
  shown by
  the solid and dash-dot
  lines, while the apastron and periastron
  are shown by dashed and dotted lines, respectively. }
\label{ls5039LC}
\end{center}
\end{figure}

The background-subtracted light curves of LS\,5039 
for 3 energy bands
are shown in Figure \ref{ls5039LC} 
as functions of the binary phase $\phi={\rm frac}[(t-T_0)/P_{\rm orb}]$, where ${\rm frac}[X]$ is fractional part of $X$, and 
$T_0$ and $P_{\rm orb}$ are the best-fit values of the epoch of periastron and binary period taken from   \cite{2009ApJ...698..514A}.
The light curves grow from  minima at $\phi\approx 0.1$ (near superior conjunction, $\phi_{\rm supc}=0.046$) up to 
$\phi\approx 0.4$, then 
remain nearly flat, with short 
fluctuations, around apastron ($\phi=0.5$), show  narrow peaks at $\phi=0.6$ (before inferior conjunction; $\phi_{\rm infc}=0.67$), and secondary peaks at $\phi\approx 0.8$. 
Thus, the light curves exhibit  a flat-top maximum, encompassing the  apastron 
and inferior conjunction phases. If, instead of the ephemeris from
\citet{2009ApJ...698..514A}, 
we use the ephemeris from Casares et al.\ (2011) or Sarty et al.\ (2011) 
for folding and/or  calculating the conjunction phases,
the shifts will not exceed 0.1 in phase.

The overall structure of the light curves does not evolve noticeably with energy, not only within the \nustar\ band but also between the  \suzaku\  XIS and  \nustar\ bands (see \citealt{2009ApJ...697..592T}). It also 
does not show appreciable changes between the  \suzaku\ 
 XIS and  \nustar\ observations,
separated by 9 years.

\begin{figure*}[hbt]
  \includegraphics[width=\textwidth]{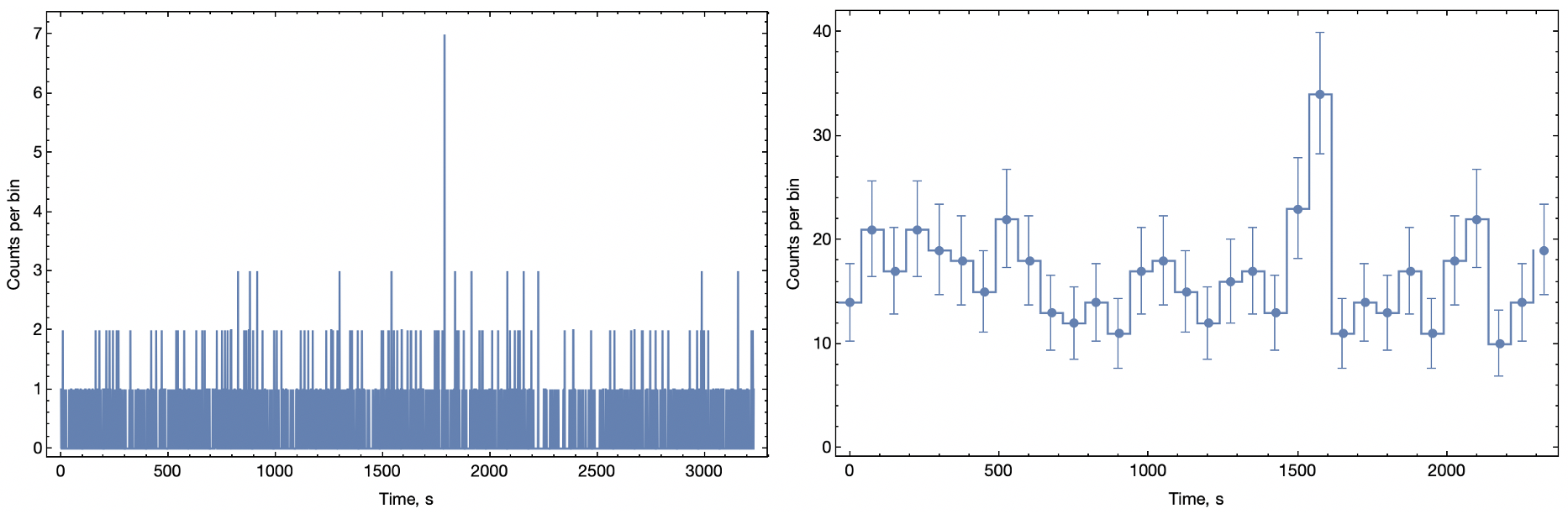}
  \caption{Light curves of the two most significant bursts (3.6$\sigma$ and 3.2$\sigma$, respectively). Bin sizes in the left and right light curves are 1 s and  75 s, respectively.
  Time 
  (in seconds)
  is counted from the end of the nearest 
  preceding occultation of the target by Earth.  }
  \label{fig:bursts}
\end{figure*}

We have also searched the data for short aperiodic variability, such as bursts. 
We 
 binned the arrival times
to produce light curves with  bin sizes varying from 1  to 200 s (so that the largest bins are smaller than the smallest gap in the \nustar\  observation). 
We calculated the Poisson probability, $q$,
of having a
 number of counts per bin larger than measured.
 Note that the average number of counts per bin (the Poisson distribution parameter) varies with the binary phase.
 This is taken into account  by calculating a  local mean over  $\approx3000$ s  
 time interval (slightly larger than the largest gap in the \nustar\  observation) surrounding the bin for which the probability is calculated (this bin itself is excluded from the mean calculation). 
 For the chosen bin size, the Poisson  probability 
 should be corrected for the number of trials, $N_{\rm tr}$,  which is 
  equal to the  number of  bins, 
 $N_{\rm bin}$, in the entire {\sl NuSTAR} observation:
  $q_{\rm corr} =1-(1-q)^{N_{\rm bin}}$.
 Figure  \ref{fig:bursts}  shows two most significant bursts that we found. The burst durations are about $\sim$1 s and $\sim$70 s.  With the above definition of  probability, the significances 
 are 3.6$\sigma$ for the short burst and 3.2$\sigma$ for the longer burst.  However, we note that this probably  
 was derived for a fixed number of bins. If we account for all possible binning schemes (not just the one that results in the highest significance), then $N_{\rm tr}>N_{\rm bin}$, i.e., the 
 confidence levels become lower.
 Therefore, these two burst candidates are not
  truly
 significant. 

\section{Spectral analysis}
\label{specana}

\subsection{Phase-integrated spectroscopy}

We 
analyzed the {\sl NuSTAR} and {\sl Suzaku}
spectra using XSPEC ver.\ 12.9.1p
\citep{arnaud96conf}.  To account for interstellar absorption, we used the T\"{u}bingen-Boulder 
model  ({\tt tbabs} in XSPEC) with the photo-electric cross-sections of
\citet{1996ADNDT..64....1V} and the abundances of \citet{2000ApJ...542..914W}.
We bin
the spectra to have 
$S/N\simeq 7$ in each spectral bin,
corresponding to 50 counts per bin, and used the $\chi^2$ statistic 
for model parameter estimation and error calculation. 
For all spectral fits, we added 
multiplicative constants to the FPMA and FPMB normalizations,
frozen to
1 for the former and allowed to vary for the latter, to account for 
calibration uncertainties between the two 
detector modules. We found 
the difference in the normalization factors not exceeding 2\%.
We also 
found, using the same approach, that the calibration uncertainty 
between the \nustar\ 
 and {\sl Suzaku}
instruments 
is around
10\%.

\src\ is detected with \nustar\ 
up to 70 keV, with a background-corrected count rate of 0.186(1) cts s$^{-1}$ and 0.171(1) cts s$^{-1}$,  in  the FPMA and FPMB detectors, respectively, in the 3--70 keV energy range,
 with a background contribution of $<10\%$ in each detector. The number of background-corrected counts in the 60--70~keV energy range is about $45\pm12$ in each
module with a  background contribution of $\approx70-80\%$, i.e.,
the source becomes 
hardly distinguishable from the background at higher energies.
An absorbed PL model gives 
a statistically acceptable fit to the phase-integrated 
 spectrum in the 3--70 keV band, with $\chi^2=667$ for
708 degrees of freedom (dof). We find a hydrogen column density
$N_{\rm H}=(0.7\pm0.4)\times10^{22}$~cm$^{-2}$, 
a photon index
$\Gamma=1.61\pm0.01$,
and an absorption-corrected  
energy flux 
$F_{\rm 3-70\,keV} =(2.31\pm0.02)\times10^{-11}$~erg~s$^{-1}$~cm$^{-2}$. There are no
indications of 
spectral features in the fit residuals,
including around
the Fe K$\alpha$ line complex 
at 6--7 keV. Moreover,
considering the \nustar\ spectra alone,
we find no evidence of a
high-energy cutoff usually present in the hard X-ray spectra of HMXBs
harboring 
accreting neutron stars 
(see, e.g., \citealt{2002ApJ...580..394C,2015ApJ...809..140K,2017ApJ...841...35F}). 
The photon index as inferred from the \nustar\
spectra is 
larger 
than $\Gamma=1.51\pm0.01$
found at 
lower X-ray energies with
\suzaku\ \citep{2009ApJ...697..592T}.

\begin{figure}[th]
\begin{center}
\includegraphics[angle=0,width=0.48\textwidth]{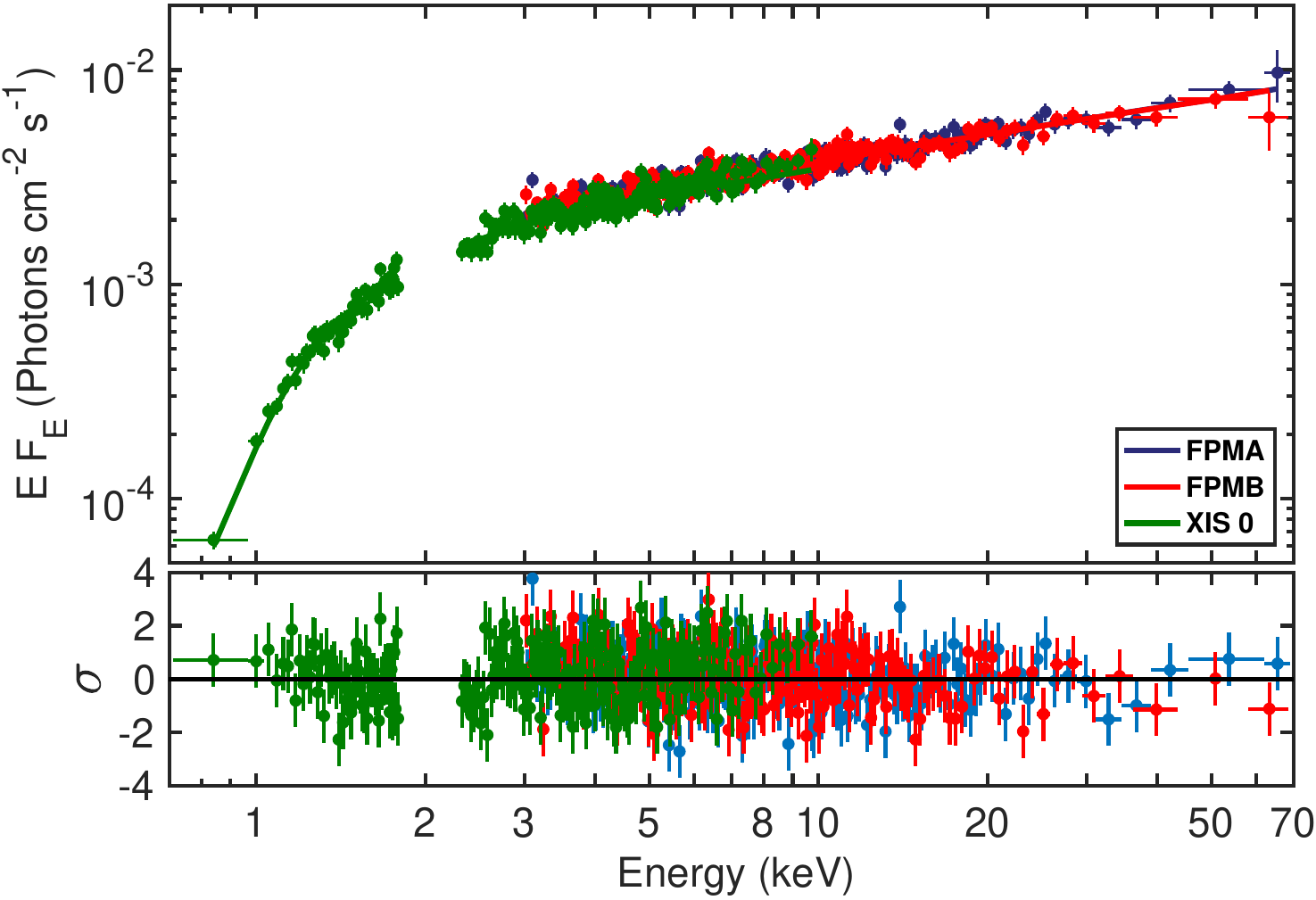}
\caption{
Fit of the absorbed PL model to the
phase-integrated \nustar\ and \suzaku\ XIS data (for clarity we only show XIS0), and the corresponding residuals in 
units of $\sigma$ (lower panel).}.
\label{suzNuSpec}
\end{center}
\end{figure}

Given the 
long-term stability
of the 
binary's periodic X-ray light curve,
\citep{2009ApJ...697L...1K}, we fit the \nustar\ 3--70~keV and the \suzaku\ 0.7--10~keV phase-averaged spectra simultaneously with an absorbed PL model (Figure~\ref{suzNuSpec}). We linked all parameters among the two spectra, except for a constant normalization to take into account  calibration uncertainties between the instruments. We find a statistically acceptable fit to the data 
($\chi^2 = 2722$ for 2795 dof), with 
$N_{\rm H}=(1.18\pm0.01)\times10^{22}$~cm$^{-2}$, 
$\Gamma=1.588\pm0.007$ (Table~\ref{specParam}),
and absorption-corrected  
fluxes
$F_{\rm 0.5-10~keV}=(9.64\pm0.05)\times10^{-12}$~erg~s$^{-1}$~cm$^{-2}$, and 
$F_{\rm
  10-70~keV}=(18.0\pm0.2)\times10^{-12}$~erg~s$^{-1}$~cm$^{-2}$.
These fluxes correspond to a luminosity $L_{\rm 0.5-70\,keV} \approx 2.8\times 10^{34} (d/2.9\,{\rm kpc})^2$~erg~s$^{-1}$.

\subsection{Phase-resolved spectroscopy}

We performed broad-band phase-resolved spectroscopy 
using the \suzaku\ XIS
and \nustar\ data
to look for 
modulation of spectral parameters  with the orbital 
period.
 We used the same definition of binary phase 
as for the 
light curve shown in Figure \ref{ls5039LC}. 
We extract the spectra in orbital phase bins $\Delta\phi = 0.1$  
($\phi=0$ corresponds to binary periastron).
Because of lower count statistics in the 
chosen phase bins, we 
bin the spectra to 5 counts per energy bin and use the Cash statistic. 
We fit all 10 spectra simultaneously with an absorbed PL. Given that phase-resolved spectroscopy with \suzaku\ alone revealed no variability in the hydrogen column density \citep{2009ApJ...697..592T}, we linked $N_{\rm H}$ between all spectra, but left the photon index and the normalization of the PL free to vary. We find a good fit to the spectra with a C-stat of 8950 for 9000 dof.

\begin{figure*}[t]
\begin{center}
  \includegraphics[angle=0,width=0.494\textwidth]{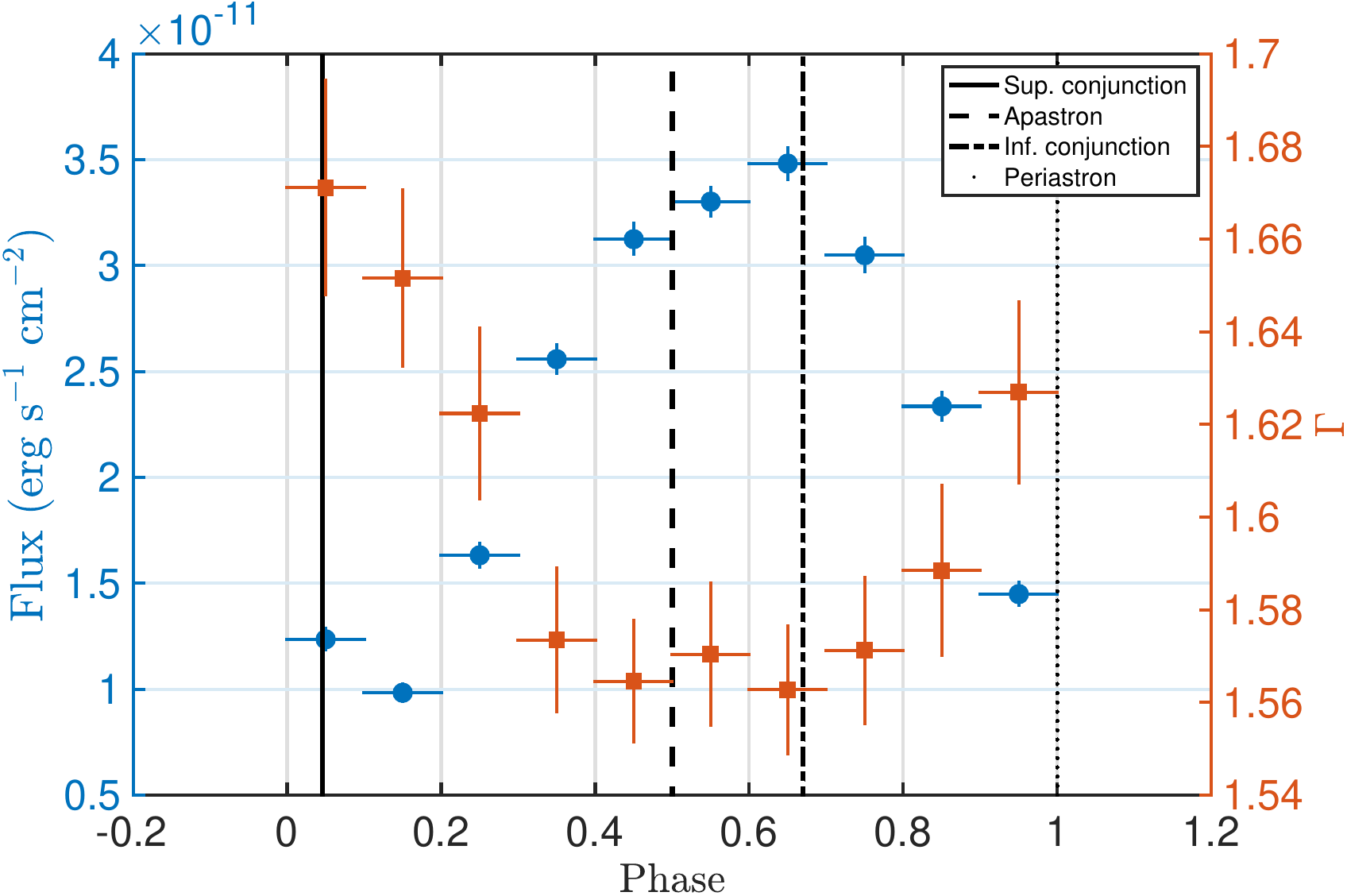}
  \includegraphics[angle=0,width=0.445\textwidth]{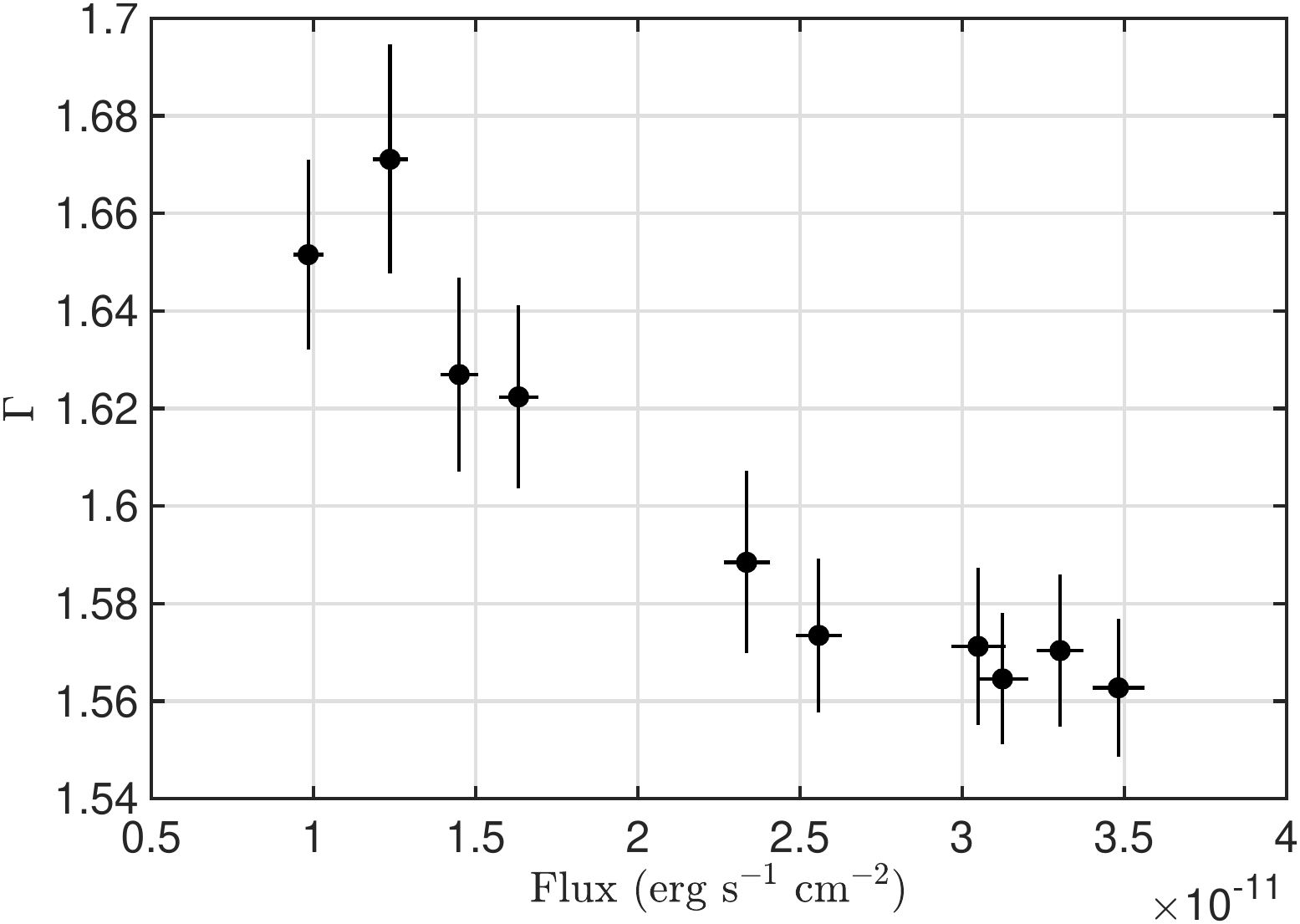}
\caption{{\sl Left:} Phase-resolved spectroscopic results with $\Delta\phi=0.1$.
  The blue points show the 3--70~keV unabsorbed flux variation as a function of
  orbital phase, with periastron at $\phi=0$. The orange squares show the photon index variation with
  phase. {\sl Right}: Anti-correlation of the  photon index 
  and the flux.} 
\label{PhResSpec}
\end{center}
\end{figure*}

The fit results are presented in Table \ref{specParam}.
Figure~\ref{PhResSpec} shows the flux and photon index 
variations 
with  orbital phase. We find a strong modulation of the photon index $\Gamma$,
as similar to those inferred previously 
from {\sl RXTE} and \suzaku\ observations \citep{2005ApJ...628..388B, 2009ApJ...697..592T}. Our \nustar\ plus \suzaku\ fits have shown harder spectra than {\sl RXTE} at all phases.
The addition of the \nustar\ data 
to the \suzaku\ data has 
provided tighter constraints 
on the photon index, showing that it varies by $\Delta\Gamma\approx0.1$ 
from maximum to minimum.

A similar tendency was noticed previously in the {\sl RXTE} phase-resolved spectra (3--30 keV band), but the values of $\Gamma$ and $\Delta\Gamma$ were substantially larger (see Figure 4 in \citealt{2005ApJ...628..388B}). A similar $\Gamma$-$F$ anti-correlation in the 1--10 keV range ({\sl XMM-Newton}, {\sl Chandra}, {\sl ASCA} and {\sl Suzaku} data) 
is shown 
by Figure  1 in \cite{2009ApJ...697L...1K}.

\begin{deluxetable}{lccc}[t]
\tablecaption{Spectral parameters of LS\,5039 from joint PL fits to the \nustar\ and {\sl Suzaku} data} \label{specParam}
\tablewidth{0pt}
\tablehead{
\colhead{Phases}
& \colhead{$N_{\rm H}$} & \colhead{$\Gamma$} &  \colhead{Flux (3-70~keV)} \\
\colhead{} & \colhead{$10^{22}$\,cm$^{-2}$}  &   \colhead{}              & \colhead{ $10^{-11}$\,erg\,s$^{-1}$\,cm$^{-2}$}
}
\startdata 
$0.0-1.0$ 
& $1.18\pm0.01$ & $1.588\pm0.007$ &  $2.31\pm0.02$\\
\hline
$0.0-0.1$  & $0.5\pm0.4$ & $1.67\pm0.02$ & $1.24\pm0.03$\\
$0.1-0.2$      & \ldots            & $1.65\pm0.02$ & $0.98\pm0.03$\\
$0.2-0.3$      & \ldots            & $1.62\pm0.02$ &  $1.63\pm0.04$\\
$0.3-0.4$      & \ldots            & $1.57\pm0.02$ &  $2.56\pm0.06$\\
$0.4-0.5$      & \ldots            & $1.56\pm0.01$ &  $3.12\pm0.06$\\
$0.5-0.6$      & \ldots            & $1.57\pm0.02$ & $3.30\pm0.05$\\
$0.6-0.7$      & \ldots            & $1.56\pm0.01$ &  $3.48\pm0.06$\\
$0.7-0.8$      & \ldots            & $1.57\pm0.02$ &  $3.05\pm0.07$\\
$0.8-0.9$      & \ldots            & $1.59\pm0.02$ &  $2.33\pm0.05$\\
$0.9-1.0$  & \ldots            & $1.62\pm0.02$ &  $1.45\pm0.04$\\
\enddata
\end{deluxetable}

\section{  Discussion and Summary
}
\label{summ}

The nature of the compact object in LS~5039 has long been elusive. 
The  two main scenarios 
  that are often used for the interpretation of its  observed properties
  are the {\em microquasar scenario} (radiatively-inefficient accretion onto a BH with the nonthermal emission coming from the jets; e.g., \citealt{2005ApJ...628..388B}), 
 and the {\em colliding winds scenario} 
 (the relativistic wind from a young rotation-powered pulsar 
 collides with the massive star's wind resulting in an intrabinary shock and particle acceleration (\citealt{2006A&A...456..801D, 2015CRPhy..16..661D, 2020A&A...641A..84M}).
 In addition, one could also consider 
 the `{\em propeller 
 scenario}' in which
 the interaction of 
 the massive star's wind with the rotating magnetosphere of a strongly magnetized NS 
 can accelerate wind particles to relativistic energies via shocks or magnetic field reconnection.
 Such a scenario was suggested by
 \cite{2012ApJ...744..106T} 
 for another HMGB, LS\,I\,+61$^\circ$\,303, from which a magnetar-like burst was likely observed, and it was also mentioned by Y+20 as a possibility for LS\,5039 if the proposed magnetar 
 nature of the compact object  is confirmed.
 The possibility of accretion onto the NS surface (magnetic poles) 
 seems to be excluded due to the long-term stability of the orbital light curve in X-rays, featureless PL spectrum from 0.3 to $\sim$100 keV, low X-ray luminosity,  $L_{X}\approx2.8\times 10^{34}(d/2.9~{\rm kpc})^2$ erg s$^{-1}$, and a lack of  outbursts.

The detection of 
fast pulsations, typical for a young rotation-powered pulsar, would 
strongly support the colliding wind scenario while 
detection of  slow pulsations 
(e.g., with a few seconds period) could support the 
scenario with a magnetar propeller, as suggested by Y+20. However, our timing analysis did not yield any evidence for a young pulsar. Although we were able to confirm 
the Fourier power excess 
near $P=9.05$ s, previously reported by Y+20,
and find a binary ephemeris 
compatible with the optical observations
(contrary to Y+20), the significance of this excess 
is rather low.
The excess near $P=8.96$ s found by Y+20 in the {\sl Suzaku} HXD data 
obtained 9 years earlier could strengthen the magnetar hypothesis, if a large enough $\dot{P}$ is assumed. However, we failed to find a 
common binary ephemeris which would 
provide even marginally acceptable signal detections at both  9.05 s (in \nustar) and 8.96 s (in {\sl Suzaku}).  Another  \nustar\  observation is needed to decisively confirm or rule out the 9 s period candidate and magnetar scenario  or, if this period is not confirmed, to perform a more sensitive search for pulsations at other frequencies. 

Our variability analysis  does not support  the accretion scenario. The binary light curve appears to be rather  stable over timescales of at least 9 years (between the {\sl Suzaku} and \nustar\ observations), including even the fine structure 
(e.g, the narrow spike near the aparstron and inferior conjuction). The stable fine structure cannot be explained by 
a clumpy stellar wind or instabilities in the accretion flow.  Understanding the nature of this stable small-scale structure in the light curve may hold the key to  the interaction scenario.

We also looked for bursts on even shorter time scales ($<200$ s), which could be expected 
in the accretion and magnetar scenarios, but found no significant bursting activity. The binary  light curve 
also shows little-to-no dependence on energy in the 0.7--70 keV range. Having the light curve maximum around the apastron (which is close to the inferior conjunction in LS 5039) phase is hardly compatible with 
wind accretion (unless the wind has very unusual properties).

The spectral analysis shows that both the phase-integrated and phase-resolved spectra in the 0.7--70 keV range  can be fitted by a simple PL model modified by the interstellar extinction. We do not find any evidence of a cutoff at higher energies. The photon index values ($\Gamma\simeq1.6$ in the phase-integrated spectrum, $1.56\lesssim\Gamma\lesssim 1.67$ in separate phase bins) are typical for such synchrotron sources as  pulsars and PWNe  \citep{2008AIPC..983..171K}. The value of $\Gamma$ is also typical for other HMGBs (see e.g., Figure~1 from \citealt{2014AN....335..301K}) but the spread of $\Gamma$ with the orbital phase is smaller for LS 5039 than those for the other HMGBs,
perhaps due to the LS 5039's tighter orbit. 

We 
have confirmed  a statistically significant anti-correlation between the flux and the photon index observed throughout the binary orbit, 
previously reported by \cite{2005ApJ...628..388B},
\cite{2009ApJ...697..592T}, and \cite{2009ApJ...697L...1K}. This phenomenon  may hold another important clue to the nature of the compact object and emission processes in LS\,5039 and the other HMGBs. A similar anti-correlation has been seen in  at least two other HMGBs: 1FGL J1018.6--5856 
\citep{2013ApJ...775..135A} 
and LS\,I\,+61$^{\circ}$303 
\citep{2017MNRAS.468.3689M}.
In the colliding wind scenario, the anti-correlation could be explained by a model in which the observer sees a harder spectrum of emitted electrons at the phases when the flux is increased by, e.g., the Doppler boosting. 
We see from Figures \ref{PhResSpec} (and \ref{ls5039LC}) that the flux is maximal and the photon index minimal near inferior conjunction, when the compact object is between the observer and the massive star. This suggests that the Doppler boosting occurs in a shocked pulsar wind confined by the dynamically dominant stellar wind in a (hollow) cone (a paraboloid-like shell)  around the star-pulsar direction (see, e.g., left panel in Figures 10  in \citealt{2013A&ARv..21...64D}).
This assumption is supported by
simulations by \cite{2012MNRAS.419.3426B} and \cite{2019MNRAS.490.3601B} 
who show that the bulk flow of the shocked pulsar wind 
can reach a bulk Lorentz factor of a few as the wind is streaming away from the  cone apex.
In order to explain the anti-correlation between the flux and photon index,
one has to assume that particle acceleration in the shocked pulsar wind proceeds more efficiently as the bulk flow accelerates, despite the adiabatic and cooling losses. The  mechanism of particle acceleration is unclear but it might be akin to that in extended AGN jets (e.g., the shear acceleration in an expanding flow; \citealt{2016ApJ...833...34R}). 

The hollow cone configuration of the shocked pulsar wind would naturally give rise to a double peaked structure in the light curve (e.g., the peaks at phases 0.4 and 0.8 in Figure \ref{ls5039LC}), as 
cone crosses the observers line of sight. This would also naturally explain the relatively flat top in the light curve. However in this scenario it is unclear what could cause the third (and strongest) peak, near phase 0.6.

Overall, we conclude that the colliding wind scenario with the compact object being a young pulsar remains the most plausible option as it can explain the dependence of the flux on the binary phase (by Doppler beaming), the spectrum (by synchrotron emission from particles accelerated by the colliding winds), the longer term stability of the light curve, and the lack of variability on short timescales.

{\em Facilities:} \facility{{\sl NuSTAR} },  \facility{{\sl Suzaku} (XRT)}

\begin{acknowledgements}

Support for this work was provided by the National Aeronautics and
Space Administration through  the \nustar\ award NNX17AB77G. JH would like to thank John Tomsick for helpful discussions regarding the reduction of Suzaku data.  JH acknowledges support from
an appointment to the NASA Postdoctoral Program at the
Goddard Space Flight Center, administered by the Universities Space Research Association under contract with NASA. GY acknowledges support from NASA through Fermi grant 80NSSC20K1571.

\end{acknowledgements}

\appendix

\section{Fourier Power Spectra and Periodic Signal Significance}
\label{appendixA}
To compute the Fourier power spectrum 
for the entire \nustar\ observation 
($T_{\rm obs}= 345,498$ s)
up to 
$f_\text{max}=1000$ Hz
using the Fast Fourier Transform (FFT), 
 we first binned the data into $T_{\rm obs}/\Delta t \approx 6.91\times10^{8}$ time bins with a bin width $\Delta t=(2f_\text{max})^{-1} = 0.0005$ s.  
 In order 
 to 
 improve the native frequency resolution ($\Delta f=T_{\rm obs}^{-1}\approx 2.89\times10^{-6}$ Hz) by approximately a factor of 5 (limited by the available computer memory size), we artificially extended the observation time span  
by that factor and assigned the average value from the real time bins to the added time bins (so that in the resulting Fourier spectrum only the DC component is affected; \citealt{2002AJ....124.1788R}). As a result, we obtained $N_\text{bin}\approx3.46\times10^9$. This corresponds to a frequency resolution  $\Delta f\approx3\times10^{-7}$ Hz.

The discrete Fourier power spectrum
is given by the following equations,
\begin{equation}
{\cal P}_n = |s_n|^2,\quad\quad\quad
 	s_n=\sqrt\frac{2}{N}
 	\sum_{k=0}^{N_\text{bin}-1}a_k \exp\left(-\frac{2\pi ikn}{N_\text{bin}}\right)\,,
 	\label{eq:fur}
 \end{equation}
 where 
  $n$ is the frequency bin number,  $a_k$ is the number of 
 counts
 in the $k$-th time bin,
 and $N=56,647$ is the total number of counts
 (in the 3--20 keV energy range). The corresponding Fourier power spectrum (for times of arrival corrected for the R\"omer delay with the best-fit orbital parameter of LS\,5039) is shown in the upper left panel of Figure \ref{fig:fur}, where we only plot 24,448 power values exceeding  ${\cal P}_n=20$. 
 
For Poisson noise,  the power in $n$-th frequency bin 
is distributed according to the exponential distribution
\begin{equation}
\varphi({\cal P}_n)
=\frac{1}{2} \exp\left(-\frac{{\cal P}_n}
{2}\right),
\end{equation}
with the mean value ${\overline{\cal P}_n} = 2$ (i.e., the same 
mean as for the 
often-used $Z_1^2$ statistic; Buccheri et al.\ 1983). 
 The probability of obtaining,
 in the Poisson noise, at least one value of power greater than ${\cal P}_n$ in
 $N_{\rm tr}=T_{\rm obs} f_{\rm max}$  independent trials
 is
\begin{equation}
\text{Pr}({\cal P}>{\cal P}_n) = 1-{\cal S}_n, \quad\quad\quad
{\cal S}_n = \left[\int_0^{{\cal P}_n} \varphi(x) dx \right]^{N_{\rm tr}} = \left[1-\exp\left(-\frac{{\cal P}_n}{2}\right)\right]^{N_{\rm tr}}
\end{equation}
The value of ${\cal S}_n$ determines the likelihood of the hypothesis that
the power ${\cal P}_n$ is generated by a periodic signal with frequency $n\,\Delta f$ rather than by the Poisson noise.

 Following tradition, we chose to characterize the $n$-th power value by the parameter  
 \begin{equation}
     \alpha_n =\sqrt{2}\, \text{erf}^{-1} {\cal S}_n\,,
 \end{equation}
 where ${\rm erf}^{-1}$ is the inverse of the error function, ${\rm erf}(x)=2\pi^{-1/2}\int_{0}^{x}e^{-z^2}dz$. 
 The value of $\alpha_n$ for a  peak in the Fourier power spectrum determines the significance 
 ``in units of $\sigma$'' of possible detection of a periodic signal with the frequency $n\,\Delta f$. 
 
 The top panel of Figure \ref{fig:fur} does not show highly significant peaks, except for those 
 caused by the orbital motion of the \nustar\ satellite (at the frequency $1.72\times 10^{-4}$ Hz and its harmonics).
 The visual impression of growing  
 ${\cal P}_n$ 
 toward higher frequencies is due to the logarithmic scale -- the number of (statistically independent) frequency bins per given log-frequency interval grows with $f$. The lack of significant detection may be caused by a spread of the signal over several frequency bins due to inaccurate values of the binary parameters
 (see Appendix \ref{appendixB}).

In addition to  the Fourier spectrum 
for the entire \nustar\ observation, we also calculated 
60 power 
 spectra for 
 separate (consecutive)
 \nustar\ orbits 
 (i.e., a dynamical 2D Fourier power spectrum), with average values of 3154 s for the visibility interval,
 $\Delta f =3.4\times 10^{-4}$ Hz for frequency resolution, and $N=944$ for
 the number of events 
 per interval.
The R\"omer delay correction was applied before calculating the spectra.
 Although the individual spectra contain less counts than the spectrum for the entire observation, they  are less prone to the loss of 
 coherence caused by  the uncertainties of orbital parameters. 
 If signal frequencies are found in each of the 60 time intervals, their dependence on interval number (time) would allow one to measure the time-dependent Doppler shift due to the difference between the best-fit and actual binary parameters, and to additionally constrain the binary parameters. 
 
 Similar to the 1D spectrum for the entire observation, 
 the Fourier power spectrum of white noise appears to increase toward higher frequencies
 when plotted on a logarithmic frequency scale.
To alleviate this issue, 
we split the frequency range into 
 $K$  segments 
of equal size on the logarithmic scale. We chose $K=200$ for visual clarity. The $k$-th ($k=1,\dots,K$) segment  contains   $n_k$  original
frequency bins, and 
this number grows with $k$. Within each frequency segment we identified the largest value of Fourier  power, ${\cal P}_k^{\rm max}$ (shown in Figure \ref{fig:fur}, bottom left). 
Similar to the 1D power spectrum, the probability of getting at least one value ${\cal P}>{\cal P}_k^{\rm max}$ within segment $k$ is 
\begin{equation}
\text{Pr}({\cal P}>{\cal P}_k^{\rm max}) = 1-{\cal S}_k, \quad\quad\quad
{\cal S}_k = \left[\int_0^{{\cal P}_k^{\rm max}} \varphi(x) dx \right]^{n_k} = \left[1-\exp\left(-\frac{{\cal P}_k^{\rm max}}{2}\right)\right]^{n_k}
\end{equation}
Similar to the above, 
we can characterize the $k$-th frequency segment by the parameter $\alpha_k$
(significance 
in units of standard deviation): 
\begin{equation}
\alpha_k=\sqrt{2}\,{\rm erf}^{-1} {\cal S}_k\,.
\end{equation}

The time-frequency image of $\alpha_k$ is shown in Figure \ref{fig:fur}  (bottom right); it is independent of the number of bins in the $k$-th frequency segment. Note that $\alpha_k$ only accounts for the number of trials within the $k$-th segment but not for the total number of trials available (independent frequency values).

\section{
Searching for a periodic signal from a binary companion when binary parameters are poorly known} 
\label{appendixB}

There can be two approaches to searching for a periodic signal from a companion in a binary system
when the binary parameters are not 
certain enough.
In the first approach one creates a sufficiently dense grid in the multi-dimensional space of binary parameters
and 
 calculates the Fourier power spectrum for the entire time series of photon arrival times after correcting them  for the R\"omer delay for all parameter values on the grid. This approach relies on the assumption that, 
 for the
 correct orbital parameter values, 
 the coherence of the periodic signal is maintained throughout the entire observation.
 Although straightforward, this approach is extremely expensive computationally if the uncertainties of the binary 
 parameters are as large as they are in the case of LS 5039 
 \citep{2012MNRAS.427.2251C}.
Therefore, we have to resort to a different approach where the observation is divided into time segments and  the coherence is maintained only locally within the each time segment. 

\begin{figure}
  \includegraphics[width=0.5\textwidth]{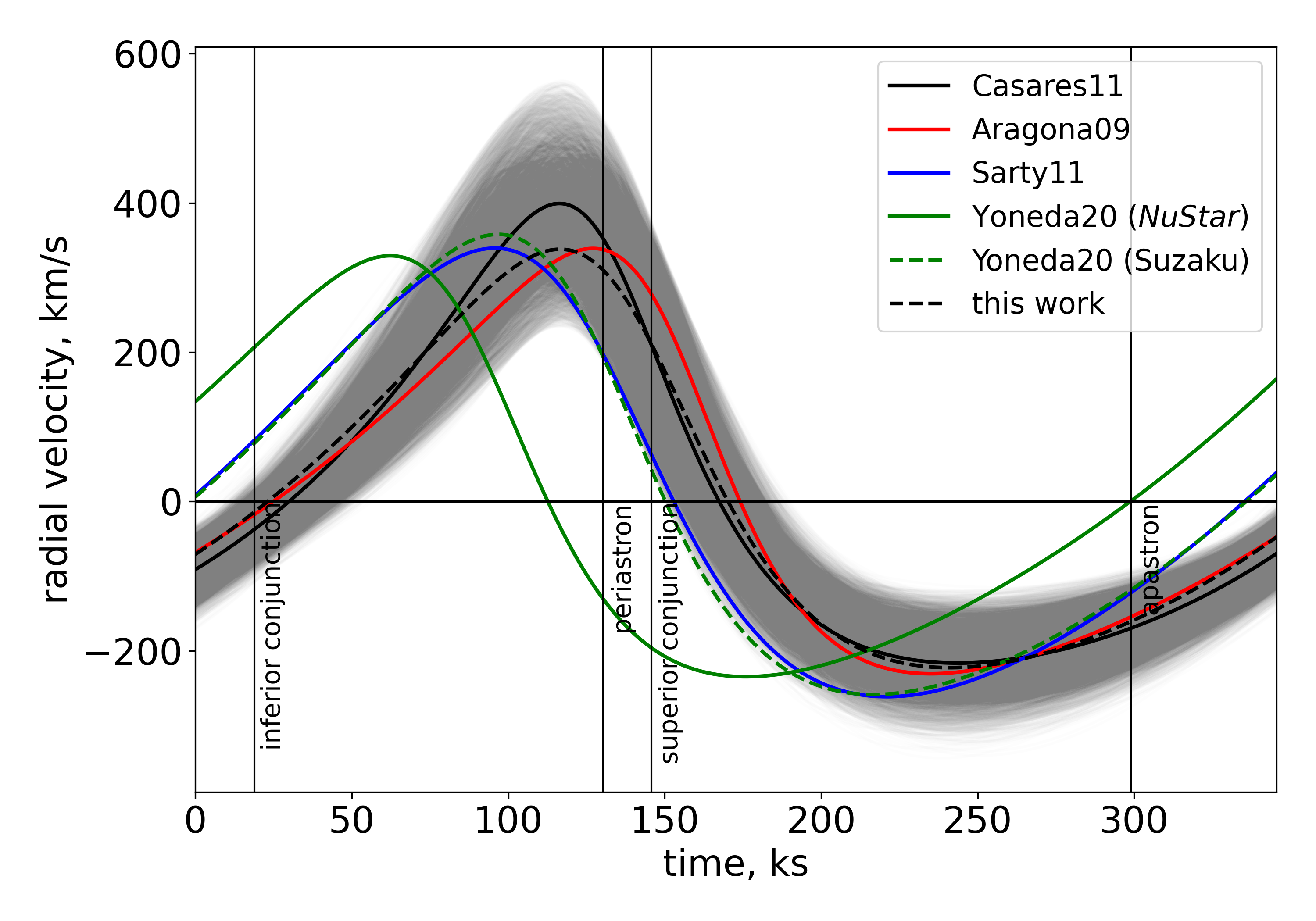}
  \includegraphics[width=0.5\textwidth]{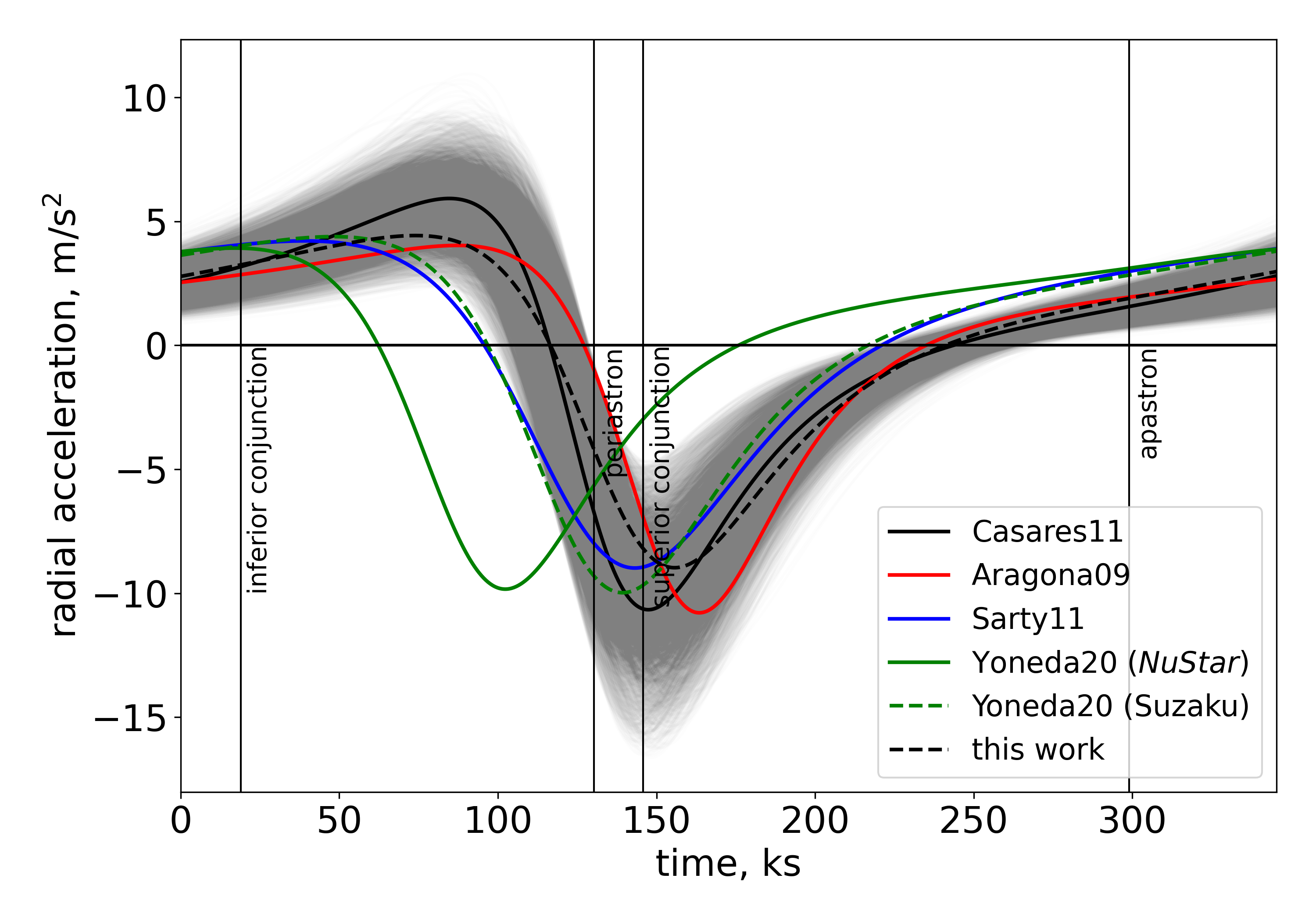}
  \caption{
  Radial projections of the velocity and acceleration of the compact object during our {\sl NuSTAR} observation 
  for 
  orbital parameters from \cite{2009ApJ...698..514A}, \cite{Casares11}, \cite{Sarty11},  Y+20, and from this work, listed in Table \ref{tab:orb}. The former 3 curves (black, red and blue) correspond to the optically 
  measured best-fit binary parameters, assuming the compact object mass of 
 $1.8 M_\odot$ for \cite{Casares11} and \cite{Sarty11}, and $1.6 M_\odot$ for \cite{2009ApJ...698..514A}. 
The solid and dashed green curves correspond to the orbital 
parameters for which Y+20 claimed the periods $P\approx 9.05$ s and $\approx8.96$ s in the \nustar\ and \suzaku\ data, respectively.
The black dashed curves correspond to the 
orbital parameters for which we found the most significant candidate for the $P\approx9.05$ s period.
The shaded areas around the solid black curves demonstrate the uncertainties of radial velocity and acceleration derived from the \cite{Casares11} parameters. 
}
  \label{fig:vel_acc}
\end{figure}

In the Appendix \ref{appendixA} and Section \ref{fourier_spec_sep_orbs}, we described the case 
when the lengths of time segments 
coincide with the 
visibility intervals in 60 separate \nustar\ orbits. However, it is not necessarily the best possible choice.  To choose the optimal number and lengths of time segments, one should take into account the frequency variation due to the Doppler effect.

If we do not apply the R\"omer delay correction, the observed frequency 
varies along the orbit as
$f(t)=f_0 [1+v_\parallel(t)/c]$,
where 
$f_0$ is the 
frequency in the reference frame of the source, and 
$v_\parallel(t)$ is the
radial
velocity of the source of signal  (see Figure \ref{fig:vel_acc}),
\begin{equation}
v_\parallel(t)=\Omega_{\rm orb}\frac{a_p\sin i}{\sqrt{1-e^2}}\left[\cos\left(\theta + W_p\right)+e\cos W_p\right],
	\label{eq:velpar}
\end{equation}
$\theta$ is the true anomaly
\begin{equation}
	\theta=2\arctan\left(\sqrt\frac{1+e}{1-e}\tan\frac{E}{2}\right)\,.   
\label{eq:trueanom}
\end{equation}

Let us consider the case when the entire observation is split into $N_w$ equal time segments (windows) of duration $T_w$ (i.e., $T_w=T_{\rm obs}/N_w$), numbered as $j=1,2,\ldots,N_w$.
If $T_w\ll P_{\rm orb}$
(which implies $N_w\gg 1$ for our case, when $T_{\rm obs}\approx P_{\rm orb}$), the radial velocity 
within the $j$-th time segment,
$t_j -T_w/2 < t < t_j+T_w/2$,
varies as $v_\parallel (t) \approx 
v_\parallel (t_j) + \dot{v}_\parallel (t_j) (t-t_j)$,
where $\dot{v}_\parallel(t_j)$ is the radial acceleration, 
\begin{equation}
	\dot{v}_\parallel(t) =-\Omega_{\rm orb}^2\frac{a_p \sin i}{(1-e^2)^2}\left(1+e\cos \theta\right)^2\sin\left(\theta + W_p\right)\,,
\label{accelpar}
\end{equation}
in the middle of the segment.
Thus, the
 frequency 
 within the $j$-th time segment
varies as 
$f_j(t) \approx f_0 [1 + v_\parallel(t_j)/c + 
\dot{v}_\parallel(t_j)\, (t-t_j)/c]$,
where the second term  corresponds to the 
frequency shift 
in the middle of the $j$-th segment, and 
the third term describes the frequency drift 
within this segment.
As $N_w\gg 1$, the maximum
drift during the entire segment,
$|\dot{v}_\parallel(t_j)| T_w f_0/c$, 
is much smaller than the 
maximal frequency 
spread, $(v_{\parallel\rm max}-v_{\parallel\rm min}) f_0/c$,  during the entire observation.

If the orbital parameters were exactly known, then the R\"omer delay correction would eliminate 
the frequency spread. However, when the 
uncertainties are large (as in the case of LS\,5039), a residual spread remains even after the 
correction, which lowers the  sensitivity of the pulsation search 
if the time segments are too wide. 
In the case of $N_w\gg1$ this spread is determined by the uncertainty of the radial acceleration.

In order to 
determine this uncertainty,
we created 
$N_{\rm orb}=1000$
orbits by random sampling from the multivariate Gaussian probability distribution of the orbital parameters with means and standard deviations 
listed in Table \ref{tab:orb}.
 For each orbit 
  we 
  calculated the mean-square deviation of the radial acceleration
  $\dot{v}_{\parallel,i}(t)$ from the most likely  
  value, $\dot{v}_{\parallel,0}(t)$, corresponding to the orbit with the best-fit orbital parameters  (Table \ref{tab:orb}):
\begin{equation}
    d_i=\frac{1}{T_{\rm obs}}\int_0^{T_{\rm obs}}\left[\dot{v}_{\parallel,i}(t)-\dot{v}_{\parallel,0}(t)\right]^2\,dt.
\end{equation}
 We then constructed a probability distribution function of the deviations, $p_{ d,i}$, 
 which appear to be log-normally distributed.
 Then 
the 
 radial acceleration uncertainty (standard deviation)
can be estimated  as
\begin{equation}
    \delta \dot{v}_\parallel(t)=
    \left\{\sum_{i=1}^{N_{\rm orb}}[\dot{v}_{\parallel,i}(t)-\dot{v}_{\parallel,0}(t)]^2 p_{d,i}\right\}^{1/2}.
\label{eq:accel_uncert}
\end{equation}
The uncertainties of radial velocity and R\"omer delay were estimated in a similar manner.
The uncertainties are shown as shaded areas in Figures
\ref{fig:deltat} and \ref{fig:vel_acc}.

Since the radial acceleration uncertainty does not change substantially  throughout the orbit, we can replace 
$\delta\dot v_\parallel(t_j)$
by an `effective 
radial acceleration' $a_{\rm eff}$.
The frequency spread caused by the radial acceleration uncertainty, $\sim f\, a_{\rm eff}\,T_w/c$, is equal to the width of the frequency bin, $T_w^{-1}$, in the Fourier spectrum if 
\begin{equation}
    N_w \sim T_{\rm obs}
    \left(f a_{\rm eff}/c\right)^{1/2}\,.
    \label{eq:optimal N_w}
\end{equation}
 Dividing into  more segments 
 would reduce  the signal strength 
 because 
 the Fourier power is proportional to the number of events and hence to $T_{w}$. Therefore, this $ N_w$ represents 
 the optimal split into time segments of equal lengths.

\begin{figure*}
  \includegraphics[width=1.0\textwidth]{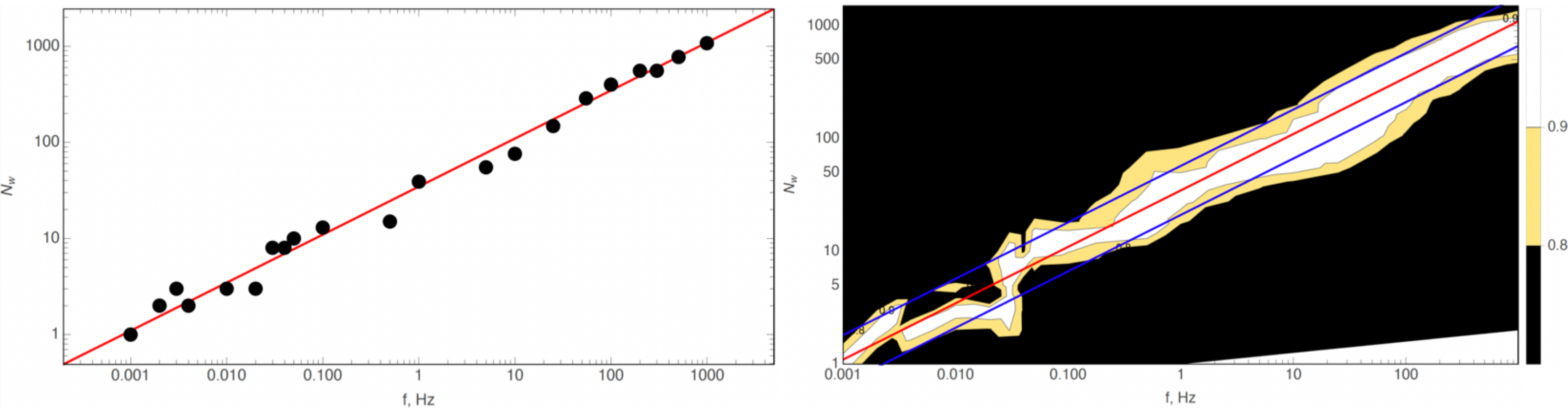}
  \vspace{-0.0cm}
  \caption{Optimal number of time segments $N_w$ vs.\ frequency $f$. Left panel: Dots 
  show the results of MC simulations; 
  the red line is a fit $N_w=34.66 f^{1/2}$ (which corresponds to effective acceleration 
  uncertainty
  of 3\,m s$^{-2}$; cf.\ Figure \ref{fig:vel_acc}, right). Right panel: Contour plot showing the areas where the deviations from optimal $N_w$ lead to the signal power being reduced by 10\% and 20\%  (shown by color). We chose such $N_w$ values so that for each frequency the signal power lies between the two blue lines.}
  \label{fig:Nw_vs_f}
\end{figure*}

 In order to 
evaluate $a_{\rm eff}$ 
for our data, 
we performed a Monte-Carlo search by generating periodic signals at various frequencies, applying a partial R\"omer correction, and finding such $N_w$ that maximizes the signal's power in a time segment (see Figure \ref{fig:Nw_vs_f}). 
We found that the dependence of $N_w$ on frequency
follows
Equation (\ref{eq:optimal N_w}) 
most closely for $a_{\rm eff}\approx  3$ m s$^{-2}$, 
comparable with the acceleration uncertainty (see Figure \ref{fig:vel_acc}, right panel), i.e., the optimal number and lengths of time segments can be estimated as $N_w\sim 35 f^{1/2}$ and $T_w\sim 9.9 f^{-1/2}$ ks in our case.

For the actual period search, it is convenient to specify a few frequency intervals with different  optimal $N_w$ numbers for each of them.
Based on the simulations, we selected  7 different sets of subdivisions of the entire $T_{\rm obs}$ into smaller 
segments, such
     that for any frequency $f$ the signal deteriorates by at most 10\%: $N_w=1$ for $f<0.017$\,Hz, $N_w=7$ for $0.017\leqslant f<0.124$, $N_w=20$  for $0.124\leqslant f<0.912$, $N_w=55$ for $0.913\leqslant f<6.74$, $N_w=148$ for $6.74\leqslant f<49.8$, $N_w=403$ for $49.8\leqslant f<368$, and $N_w=854$ for $f\geqslant368$\,Hz.

\begin{figure}[bt]
\includegraphics[width=\textwidth]{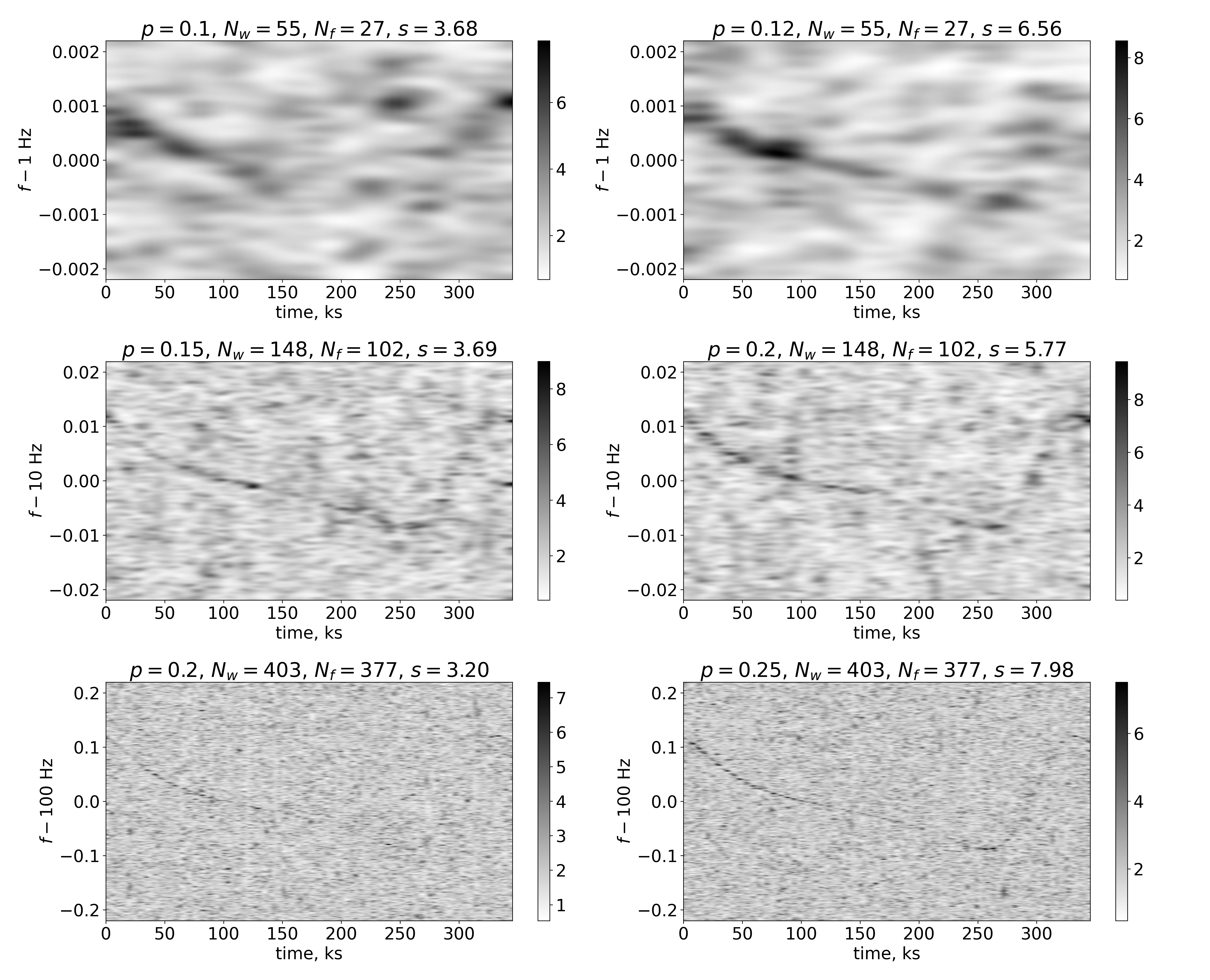}
\caption{Images in the time-frequency plane of the Fourier power 
  for artificial periodic  signals after imperfect (due to the imprecisely known emphemeris) R\"omer delay correction for 
3 signal frequencies 
and several
    pulsed fractions $p$.
Each image has $N_w\times 5N_f$ `pixels', where $N_w$ is the number of time (orbital phase) windows, and $N_f$ is the number of frequency 
segments 
(we used the frequency step $(5T_w)^{-1}$ for greater  sensitivity and visual clarity).
We have applied a Gaussian blur to make the periodic signal visible by eye.
Color bars 
indicate the magnitude of the Fourier power 
    ${\cal P}_n$,
     defined in Equation (\ref{eq:fur}).
On top of each plot we show the corresponding value of the $s$-statistic, defined in Section \ref{dynamicfourier}. 
In all the simulations we assumed that the binary has the following `true' orbital parameters: \{$a_p\sin i=51.7$ lt-s, $e=0.35$, $T_0=52477.58$ MJD, $P_\text{orb}=3.90608$ d, and $W_p=32^\circ$\} but applied the R\"oemer delay correction using a different set: \{$a_p\sin i=45.2$ lt-s, $e=0.33$, $T_0=52477.55$ MJD, $P_\text{orb}=3.90604$ d, and $W_p=29.5^\circ$\}. 
}
\label{fig:mc2}
\end{figure}

 Examples of simulated 
 dynamic Fourier power spectra in the time-frequency domain (time in units of orbital phase) are shown in Figure \ref{fig:mc2}
 for sinusoidal pulsations with the intrinsic
 frequencies $f_0=1$, 10 and 100\,Hz
 and various pulsed 
 fractions $p$
 (defined as the ratio of the number of signal events to the total number of events).

In order to detect the signal 
in the dynamic Fourier power spectrum, we split 
the entire frequency range
into segments 
that are wide enough to ensure that they contain the entire signal (i.e., the entire $f(\phi)$ curves, similar to those in Figure \ref{fig:mc2}, are within these frequency segments). 
The half-width 
of a frequency segment can be estimated as
$\delta f = \beta f_0$.
The coefficient $\beta$ is 
 proportional to $\delta v_\parallel/c$, where $\delta v_\parallel$ is the maximum 
 amplitude 
 of the residual uncertainties of  radial velocity. 
 The Monte Carlo (MC)
 simulations 
 show that 
 $\delta v_\parallel$ 
 exceeds $300$ km s$^{-1}$ 
 (i.e., 
 $\delta v_\parallel/c 
> 10^{-3}$)
 in only $\approx1 \%$ of simulations.  
To increase the chances to capture a signal in a single frequency segment,
we conservatively chose $\beta= 
2.2\times10^{-3}$. 
For a time window of width $T_w$, the number of frequency resolution bins within such a frequency segment 
around a central frequency $f_0$ is 
$N_f = 2\beta f_0 T_w$.

The dependence of the central frequency $f_m$ of a segment on its number $m$ follows from the relationship $f_m +\beta f_m = f_{m+1}-\beta f_{m+1}$ :
 \begin{equation}
 f_{m}=\left(\frac{1+\beta}{1-\beta}\right)^{m-1} f_1\,,
\label{eq:fm}
 \end{equation}
where $f_1$ is the central frequency of the first segment
within some frequency interval.
The total number of frequency segments
in the range ($f_{\rm min}$, $f_{\rm max})$ can be estimated as
\begin{equation}
    m_{\rm max}\approx 1+
    \frac{1}{2\beta} \ln\frac{f_{\rm max}}{f_{\rm min}}.
\end{equation}
For instance, $m_{\rm max} \approx 
2500$ 
for $f_{\rm max}=1000$ Hz, $f_{\rm min} = 0.017$ Hz, $\beta = 2.2\cdot 10^{-3}$.

\end{document}